\documentclass[11pt,a4paper]{article}
\pdfoutput=1
\usepackage{jheppub}
\usepackage{pdflscape}
\usepackage{bm}

\def\IZ{{\mathbb Z}}

\newcommand{\tr}{{\rm Tr}}
\newcommand{\re}{{\rm e}}
\newcommand{\ri}{{\rm i}}
\newcommand{\rd}{{\rm d}}

\newcommand{\be}{\begin{equation}}
\newcommand{\ee}{\end{equation}}
\newcommand{\ba}{\begin{aligned}}
\newcommand{\ea}{\end{aligned}}
\newcommand{\ben}{\begin{eqnarray}\displaystyle}
\newcommand{\een}{\end{eqnarray}}

\title{TBA-like integral equations from quantized mirror curves}

\author[a]{Kazumi Okuyama}
\author[b]{and Szabolcs Zakany}

\affiliation[a]{Department of Physics, Shinshu University,
Matsumoto 390-8621, Japan
}
\affiliation[b]{D\'epartement de Physique Th\'eorique, Universit\'e de Gen\`eve, Gen\`eve, CH-1211 Switzerland
}

\emailAdd{kazumi@azusa.shinshu-u.ac.jp, szabolcs.zakany@unige.ch}

\abstract{Quantizing the mirror curve of certain toric Calabi-Yau (CY) three-folds leads to a family of trace class operators. The resolvent function of these operators is known to encode topological data of the CY. In this paper, we show that in certain cases, this resolvent function satisfies a system of non-linear integral equations whose structure is very similar to the Thermodynamic Bethe Ansatz (TBA) systems. This can be used to compute spectral traces, both exactly and as a semiclassical expansion. As a main example,  we consider the system related to the quantized mirror curve of local $\mathbb P^2$. According to a recent proposal, the traces of this operator are determined by the 
refined BPS indices of the underlying CY. We use our non-linear integral equations to test that proposal. }

\begin{document}
 \maketitle
  						%
 \section{Introduction}
 
It has been found in the 90' the following curious fact \cite{CFIV,Zam}: a certain system of non-linear integral equations, coming from the Thermodynamic Bethe Ansatz (TBA) analysis of a supersymmetric index, can be solved in terms of the resolvent function of an integral operator. At first, the result was not obvious mathematically, and relied on some physical argument (in a rather convoluted way, involving the Painlev\'e III equation). The mathematical proof was put forward by Tracy and Widom in \cite{TW} a few years later, thus establishing a slight generalization of the result (already anticipated in \cite{Zam}). To quote this result:
given a ``potential'' $V(\theta)$, the solutions $\epsilon(\theta)$ and $w(\theta)$ of the system of  non-linear integral equations 
 \be
 \ba
 	\label{prototypeTBA}
 	\eta(\theta) &= 2 \kappa  \int_{-\infty}^{\infty} \rd \theta' \, \frac{\re^{-\epsilon(\theta')}V(\theta')}{\cosh(\theta-\theta')},\\
 	\epsilon(\theta) &= -\frac{1}{2\pi} \int_{-\infty}^{\infty} \rd \theta' \, \frac{\log(1+\eta^2(\theta'))}{\cosh(\theta-\theta')}, \\
	w(\theta)&=\frac{1}{\pi} \re^{-\epsilon(\theta)}\int_{-\infty}^{\infty} \rd \theta' \, \frac{\arctan \eta(\theta')}{\cosh^2(\theta-\theta')},
\ea
 \ee
 are encoded in the resolvent function of the operator $\rho$ with integral kernel
 \be
 	\label{prototypekernel}
	\rho(\theta,\theta')=\frac{\sqrt{V(\theta)V(\theta')}}{\cosh(\frac{\theta-\theta'}{2})}.
 \ee
Let us be more precise. The resolvent kernel $R(\theta,\theta')=(\rho(1-\kappa \rho)^{-1})(\theta,\theta')$ can be decomposed according to its $\mathbb{Z}_2$-parity under $\kappa\to-\kappa$ into an even part $R_+$ and odd part $R_-$. Then, the diagonals $\theta=\theta'$ of the kernels of these spectral functions are given by
\be
	R_+(\theta,\theta)=\re^{-\epsilon(\theta)}V(\theta), \qquad \qquad R_-(\theta,\theta)=w(\theta)V(\theta).
 \ee
In short, the solution of the non-linear TBA integral system (\ref{prototypeTBA}) is characterized by some spectral function of the operator (\ref{prototypekernel}). This is a quite remarkable result, which is also reminiscent of the ``ordinary differential equation - integrable model'' (ODE-IM) correspondence. Indeed, one of the results of the ODE-IM correspondence is the identification of the spectral determinant of many differential equations with the solutions of TBA-like integral equations (for a review of this subject and a list of references, see \cite{DDT}).

Of course, now that this relationship between the kernel $\rho$ and the system (\ref{prototypeTBA}) is established, we can turn the logic around, and use the TBA system to learn about the spectrum of the linear operator $\rho$. This is exactly what was done in \cite{PY,HMOexact,CM} for the kernel (\ref{prototypekernel}) with a specific family of potentials $V(\theta)$, in order to compute the partition function of the $\mathcal N=6$ Chern-Simons-matter theory known as ABJM theory through its localized matrix integral expression. Through large $N$ duality, this study allowed the investigation of the non-perturbative structure of the M-theory on $AdS_4\times 
S^7/\mathbb{Z}_k$, which is holographically dual to the $U(N)_k\times U(N)_{-k}$ ABJM theory on $S^3$.

Another, though not completely unrelated field of research where integral kernels similar to (\ref{prototypekernel}) come into play is topological strings and non-compact mirror symmetry. In a recent series of works including \cite{HMMO, KallenMarino, HW} and culminating in \cite{GHM}, it was proposed that the free-energy of both the standard (closed) topological string and its refined version in the Nekrasov-Shatashvili limit is encoded in the spectral determinant of an operator related to the mirror curve of the background geometry. This proposal is formulated for a family of non-compact geometries which are the toric Calabi-Yau (CY) threefolds. It has already passed many tests, both numeric and analytic. In order to further test the proposal for a given toric CY, one has to investigate the spectral quantities of the operator arising from the quantization of its mirror curve. Fortunately, in some cases, the integral kernel of this operator could be written down explicitly \cite{KM}. In all such cases it can be put in the form
\be
	\label{generalisedkernel}
	\rho(\theta,\theta')=\frac{\sqrt{V(\theta)V(\theta')}}{\cosh(\frac{\theta-\theta'}{2}+\ri \pi C)},
\ee
with $V(\theta)$ some real function and $C$ a rational number, which can be taken to be $|C|<\frac{1}{2}$ without loss of generality.%
\footnote{We will not consider the case where $C$ is non rational real which is expected to be a more complicated problem. Also, we want to avoid the case where $C$ is in $\IZ+ \frac{1}{2}$, because in that case the kernel becomes singular. Finally, one can use the periodicity of the $\cosh$ function to reduce any other case to $|C|<\frac{1}{2}$}
The kernel (\ref{prototypekernel}) which is related to the TBA system (\ref{prototypeTBA}), is the particular case with $C=0$ of this more general family. So it is a legitimate question to ask if the TBA system can be generalized to the case $C \neq 0$: can we build a TBA-like system whose solution is encoded in the diagonal resolvent of the kernel (\ref{generalisedkernel})? The answer is yes.
Building on the ideas that led to the proof of \cite{TW} for the prototypical kernel (\ref{prototypekernel}), we propose here a new set of TBA-like equations which are satisfied by the diagonal resolvent of the more general operator (\ref{generalisedkernel}).
Among the most important features of the construction given in \cite{TW} are the following two points: the splitting of the resolvent into the appropriate components ($R_+$ and $R_-$), and the existence of a certain bi-linear relation which we call the ``quantum Wronskian relation'' (this is the content of the Proposition of \cite{TW}).
Thus, our generalization will involve first finding the proper splitting of the resolvent function into the right building block functions, and second, properly generalizing the so called quantum Wronskian relations underlying the TBA system.
 We will mostly focus on the case $C=\frac{1}{6}$ which somehow turns out to be the second simplest case. Also, it is of particular interest because it is the kernel associated to the mirror curve of the toric Calabi-Yau given by $\mathcal O(-3) \rightarrow \mathbb P^2$, also known as
local ${\mathbb P^2}$, which is a particularly simple testing ground for the proposal of \cite{GHM}. Thus we can for example use the TBA-like equations for the local $\mathbb P^2$ kernel to compute its traces. These traces are actually predicted by the proposal of \cite{GHM} and we can compare the results as a further testing of it. The TBA system can also be studied in the semiclassical limit (we will soon introduce the $\hbar$ parameter which controls this limit), providing an efficient alternative tool to WKB methods if one wants to compute the semiclassical traces of the underlying operator. This path was taken in \cite{CM} for the family of operators relevant for ABJM theory (which have $C=0$). As a remark, it is now well known that the kernel with $C=0$ also arises from the quantization of the mirror curve of another toric Calabi-Yau  called local $\mathbb P^1 \times \mathbb P^1$, or equivalently local $\mathbb F_0$, and actually relates in a particular way to the ABJM kernel \cite{KMZ}. From the point of view of the TBA, the local $\mathbb P^1 \times \mathbb P^1$ case is then simpler than the local $\mathbb P^2$ case.

Since it is the main result of this work, let us write down the TBA-like system we found for $C=\frac{1}{6}$. As for the $C=0$ case, we can write it as a system of three non-linear integral equations for three unknown functions  $\epsilon(\theta)$,  $w(\theta)$ and $\eta(\theta)$ ($\eta$ is just an auxiliary function):
 \be
 \ba
 	\label{Cis1over6TBA}
 	\eta(\theta) &= 2 \kappa  \int_{-\infty}^{\infty} \rd \theta' \, \frac{\sqrt{3}\re^{-\epsilon(\theta')}V(\theta')}{1+2\cosh(\theta-\theta')},\\
 	\epsilon(\theta) &= -\frac{1}{2\pi} \int_{-\infty}^{\infty} \rd \theta' \, \frac{\sqrt{3}\log[(1+\eta(\theta'))(1-\eta(\theta')-\eta(\theta'-\frac{2\pi \ri}{3}))(1-\eta(\theta')-\eta(\theta'+\frac{2\pi \ri}{3}))]}{1+2\cosh(\theta-\theta')}, \\
	w(\theta)&=\frac{1}{\pi} \re^{-\epsilon(\theta)}\int_{-\infty}^{\infty} \rd \theta' \, \frac{3 \log[\re^{-\epsilon(\theta'+\frac{2\pi \ri}{3})-\epsilon(\theta'-\frac{2\pi \ri}{3})}(1+\eta(\theta'))^{-2}]}{16\cosh^2(\theta-\theta')}.
\ea
 \ee
 From the solutions $\epsilon(\theta | \kappa )$ and $w(\theta | \kappa)$ of the above $\kappa$-depending 
 system\footnote{As written in (\ref{Cis1over6TBA}), the integral system needs regularization, because when $\theta$ is replaced by $\theta+2\pi \ri/3$ in the kernel $(1+2\cosh(\theta-\theta'))^{-1}$, this kernel becomes singular, and the integration hits a pole at $\theta'=\theta$. So the shifted functions $\eta(\theta'\pm 2\pi \ri/3)$ and $\epsilon(\theta'\pm 2\pi \ri/3)$ in the second and third line of (\ref{Cis1over6TBA}) are ill-defined. A good regularization is given by replacing every $\pm 2\pi \ri/3$ by $\pm(2\pi \ri/3-\ri \delta)$ and taking the $\delta \rightarrow 0$ limit.  },
 one can reconstruct the diagonal resolvent of $\rho$ given in (\ref{generalisedkernel}) with $C=\frac{1}{6}$. In order to do that, we have to decompose the diagonal resolvent kernel $R(\theta,\theta)=(\rho(1-\kappa \rho)^{-1})(\theta,\theta)$ according to a $\mathbb Z_3$-charge under $\kappa\to
\omega\kappa$
into $R_k(\theta,\theta)$ with $k=0,1,2$,
where $\omega=\re^{2\pi \ri/3}$ is the generator of $\mathbb Z_3$. Then, we have that
 \be
 \ba
 	R_0(\theta,\theta)+R_1(\theta,\theta) &=\frac{4}{3\sqrt{3}} {\rm Re}((1-\omega^{-1}) \, \re^{-\epsilon(\theta| \omega \kappa)}) V(\theta), \\
	R_2(\theta,\theta) &=\left (\frac{1}{2}w(\theta| \kappa)+{\rm Re}(\omega \, w(\theta | \omega \kappa )) \right ) V(\theta).
\ea
 \ee
 As before, this is valid for any real ``potential'' function $V(\theta)$ which decays nicely at infinities. The appearance of the cyclic group $\IZ_3$ is a consequence of the particular value taken by $C$, and it is related to the underlying $\IZ_3$ symmetry of the quantized mirror curve of local $\mathbb P^2$
\cite{KM}.
 In general its role is played by a higher cyclic group\footnote{The fact that this underlying group is finite is a consequence of the assumption $C \in \mathbb Q$.
 }; we will see more details concerning the general case in section \ref{sec:TBA-mn}. 

The paper is organized as follows. In section \ref{sec:TBA-P2}, we write down the quantized mirror curve of local $\mathbb P^2$, recall the relevant kernel related to it, and define its resolvent. We then derive the TBA-like system for that case, and study its solution in the semiclassical limit  (as done in \cite{CM} for the kernel given in (\ref{prototypekernel})). We choose to present the detailed derivation for the special case of local $\mathbb P^2$, so that the essentials of the derivation are not hidden behind the cluttered notation.
In section  \ref{sec:exact-Z},
we will compute the exact values of the spectral traces for the local $\mathbb{P}^2$
for a couple of values of $\hbar$,
and find the perfect agreement with the conjecture in \cite{GHM}.
It turns out that all the steps in section \ref{sec:TBA-P2}
are straightforwardly generalizable to arbitrary rational  $C$ : this is what we present in section 
\ref{sec:TBA-mn}, and obtain a general TBA-like system in difference form, which can be inverted for each particular $C$ into a true integral system. We conclude in section \ref{sec:conclusion}.
In the appendices \ref{apptracerho}
and \ref{appQW} are given some finer points of the derivation in section \ref{sec:TBA-P2}, and in appendix \ref{app3} are gathered some results for local $\mathbb P^2$ when $\hbar=6\pi$.
 
%
%
%
 
  						%
 \section{The TBA system for local $\mathbb P^2$}
\label{sec:TBA-P2}
 
 \subsection{The integral kernel and the resolvent}
 Let us introduce canonically conjugate operators $\rm x, \rm y$ with $[\rm x, \rm y]=\ri \hbar$.
 The operator corresponding to the toric Calabi-Yau threefold known as local $\mathbb P^2$ can be written as \cite{GHM}
 \be
 	\rm O_{\mathbb P^2} = \re^{\rm x}+\re^{\rm y}+\re^{-\rm x -\rm y}.
 \ee
 As usual in quantum mechanics, one can represent the conjugate pair $\rm x$ and $\rm y$ as multiplication and differentiation operations to be applied on a wave function. In such representation, the operator $\rm O_{\mathbb P^2}$ corresponds to a difference operator. The inverse operator 
 \be
 	{\rho}_{\mathbb P^2} = \rm O^{-1}_{\mathbb P^2}
 \ee
 has been proven to be a trace class operator on $L^2(\mathbb R)$ and an explicit integral kernel has been found in terms of the quantum dilogarithm \cite{KM}. Introducing a new basis given by ${3\rm x}=2 {\rm p}+{\rm q}, {3\rm y}=-{\rm p}-{2 \rm q} $, with $[{\rm q},{\rm p}]= \ri h \equiv 3\ri \hbar$, the operator ${\rm \rho}_{\mathbb P^2}$ can be factorized as
  \be
 	{\rho}_{\mathbb P^2} =W({\rm p}) \frac{\re^{ \frac{1}{6}\rm q} }{2 \cosh \left ( \frac{1}{2}\rm q \right )} W({\rm p})^*,
\label{rho-W}
 \ee
 where $W({\rm p})$ is given by\footnote{For the definition of Faddeev's quantum dilogarithm $\Phi_b(z)$ and some of its properties, see for example \cite{KM}.}
 \be
 	W({\rm p})=\re^{\frac{1}{6} {\rm p}} \Phi_b\left (\frac{{\rm p} +\ri \hbar}{2 \pi b} \right ), \qquad \qquad b^2= \frac{3 \hbar}{2\pi}.
\label{Wp}
\ee
The $^*$ in \eqref{rho-W} denotes the Hermitian adjoint. It is straightforward to obtain the integral kernel in the $p$ representation from the above expression. Since we will be only interested in quantities related to the spectrum of ${\rm \rho}_{\mathbb P^2}$, we can study another operator which is related by a similarity transformation: 
  \be
  \label{P2kernel}
 	{\rm \rho}=\sqrt{V(\rm p)}  \frac{\re^{ \frac{1}{6}\rm q} }{2 \cosh \left ( \frac{1}{2}\rm q \right )}  \sqrt{V(\rm p)}, \qquad \qquad V({\rm p})=W({\rm p})W({\rm p})^*.
\ee
 As it is the case in the prototypical example given in \cite{Zam,TW}, the structure of the TBA-like integral equations will not depend on the details of the $V(p)$ function. So let us consider the more general family of kernels
 \be
	\label{rhofullkernel}
 	{\rm \rho}(p,p') =\frac{\sqrt{V(p)V(p')}}{2 h \cosh \left ( \frac{\pi}{h}(p-p')+\frac{\ri \pi}{6}\right )} = \frac{\sqrt{E(p) E(p')}}{\alpha M(p) +\alpha^{-1} M(p')},
 \ee
 where we define
  \be
  \ba
 	\alpha & =\re^{\ri \pi /6}, \qquad \qquad \qquad \qquad \omega \equiv -\alpha^{-2}=\re^{2\pi \ri /3},\\
	E(p)&=\frac{1}{h}M(p) V(p), \qquad \quad M(p) =\re^{\frac{2\pi}{h}p},
\ea
 \ee
 and where $V(p)$ is an arbitrary positive function 
 with good enough analyticity properties so that $\rho$ is of trace class. To retrieve the local $\mathbb P^2$ case, we should use the specific potential $V(p)$ given by
the second equation of \eqref{P2kernel} with $W({\rm p})$ in \eqref{Wp}. We remark that the kernel which is considered in \cite{Zam,TW} and given in (\ref{prototypekernel}) has the same structure, but with $\alpha=1$ (corresponding to $C=0$). Let us define recursively the following series of functions
  \be
  \ba
  	\label{phijdef}
 	\phi_j(p)= \int_{-\infty}^{\infty} \rd p' \frac{1}{\sqrt{E(p)}} \rho(p,p') \sqrt{E(p')} \phi_{j-1}(p'), \qquad \qquad \phi_0(p)=1.
\ea
 \ee
 Except for $\phi_0$, these are all functions in $L^2(\mathbb R)$. It is not too difficult to check (see appendix \ref{apptracerho}) that the kernel of the $\ell^{\rm th}$ power of the operator can be written as
 \be
 	\label{powerkernel}
	\rho^{\ell}(p,p')=\alpha^{-1} \frac{\sqrt{E(p)E(p')} }{M(p)-\omega^\ell M(p')}\sum_{j=0}^{\ell-1} \omega^j \phi_{j}(p) \bar \phi_{\ell-1-j}(p'),
\ee
where $\bar \phi$ denotes the complex conjugate of $\phi$. This is actually valid for arbitrary value of $\alpha$. Expressions (\ref{phijdef}) and (\ref{powerkernel}) can already be used to efficiently generate exact traces. The corresponding method was used on the $\alpha=1$ kernel in \cite{HMOexact,HMOinst} in the context of ABJM, to obtain exact traces for high $\ell$.

We now define the resolvent operator $\rm R(\kappa)$ as the formal sum
  \be
  	\label{resolvent}
	{\rm R}(\kappa)=\frac{\rho}{1-\kappa \rho}=\sum_{\ell = 0}^\infty \kappa^{\ell} \rho^{\ell+1} .	
\ee
 The trace of $\rm R(\kappa)$ is the generating function of the spectral traces  $\tr \rho^{\ell}$, and it is related to the grand potential $\mathcal J$ of \cite{GHM} as
  \be
  	\label{RtoJ}
	\tr   \, {\rm R} (-\kappa)=\frac{\rd}{\rd \kappa} \mathcal J(\kappa).
\ee
The grand potential $\mathcal J$ of \cite{GHM} is in fact the logarithm of the Fredholm determinant of the operator $\rho$ 
 \be
  	\mathcal J(\kappa)=\log \det (1+\kappa \rho),
\ee
 a fundamental function related to the spectrum of $\rho$.

In the case $\alpha=1$ ($\omega=-1$), there is a natural splitting of the resolvent into an odd and an even part \cite{Zam,TW}, which do not enter symmetrically in the TBA system. This comes from the fact that $\omega$ is a square-root of 1, the non-trivial element of $\mathbb Z_2$. In our case, we have $\omega=\re^{\frac{2 \ri \pi}{3}}$, which is a non trivial element of $\mathbb Z_3$. We thus expect that the natural splitting of the resolvent is in components with definite ``charge'' 
under the $\IZ_3$ symmetry $\kappa\to\omega\kappa$
 \be
 	\label{RkfromR}
	{\rm R}_k(\kappa)=\frac{1}{3}\sum_{r=0}^{2} \omega^{-r k} {\rm R}(\omega^r \kappa)=\frac{\kappa^k \rho^{k+1}}{1-(\kappa \rho)^3}, \qquad \qquad (k=0,1,2)
\ee
such that
\be
\ba
	\label{Rkrelations}
	{\rm R}_k(\omega \kappa)&=\omega^k {\rm R}_k(\kappa), \qquad \qquad
	{\rm R}(\kappa)=\sum_{k=0}^{2}{\rm R}_k(\kappa).
\ea
\ee
If we define $\Phi_r$ the generating functions\footnote{Not to be confused with the Faddeev quantum dilogarithm, which will not turn up in what follows.} of $\phi_j$ and its complex conjugate as
\be
\ba
	\Phi_r &= \sum_{j = 0}^\infty \kappa^{r+3j} \phi_{r+3j}, \qquad \qquad  	\bar \Phi_r = \sum_{j = 0}^\infty \kappa^{r+3j} \bar \phi_{r+3j} \qquad \qquad 	 (r=0,1,2),
\ea
\ee
we can write down the integral kernel of the ${\rm R}_k$ resolvent as
\be
\ba
	\label{Rkkernel}
	R_k(p,p')=\alpha^{-1} \frac{\sqrt{E(p)E(p')} }{M(p)-\omega^{k+1} M(p')}\sum_{r=0}^{2} \omega^r \Phi_{r}(p) \bar \Phi_{k-r}(p').
\ea
\ee
In expression (\ref{Rkkernel}) and in what follows, the indices of $\Phi_{r}$ and $\bar \Phi_{r}$ are to be taken modulo $3$. The quantities that will enter into the TBA-like integral equations are the diagonal elements of the resolvent kernels
\be
	\label{Rkdef}
	R_k(p) \equiv \lim_{p \rightarrow p'} R_k(p,p'),
\ee
which can be used to obtain the generating function of the traces as
\be
	\label{limofkResolvent}
	\int_{-\infty}^{\infty} \rd p \,  R_k(p) = \sum_{j = 0}^\infty \kappa^{k+3j} 
\text{Tr}(\rho^{k+1+3j}).
\ee
Because our operator is of trace class, we expect the $R_k(p)$ to be well defined functions with finite integral over real $p$. But the denominator of $R_{2}$ in (\ref{Rkkernel}) vanishes when $p \rightarrow p'$, which implies that the numerator should also vanish in this limit:
 \be
 	\label{zerocomb}
 	\sum_{r=0}^{2} \omega^r \Phi_{r}(p) \bar \Phi_{2-r}(p) =0.
 \ee
 Taking this into account, we end up with the following expressions for the diagonal resolvent kernels:
 \be
\ba
	\label{nonvanishingRcomb}
	R_k(p)&=\frac{V(p)}{h} \frac{1}{2\sin \left ( \pi\frac{k+1}{3} \right )}\sum_{r=0}^{2} \omega^{r-\frac{k}{2}} \Phi_{r}(p) \bar \Phi_{k-r}(p), \qquad \qquad (k = 0,1), \\
	R_{2}(p)&=-\frac{V(p)}{4\pi} \sum_{r=0}^{2}\omega^{r+\frac{1}{2}} \{\Phi_r(p),\bar \Phi_{2-r}(p) \},
\ea
\ee
where we use the notation $\{f(p),g(p)\}=\ri ( f'(p)g(p)-f(p)g'(p) )$. The expressions (\ref{nonvanishingRcomb}) are to be taken as the direct equivalents of Lemma 1 in \cite{TW}.
  
\subsection{Build up the TBA equations for local $\mathbb P^2$}

What we call the TBA equations for local $\mathbb P^2$ are a system of non-linear integral equations 
satisfied by a certain combination of the diagonal resolvent functions $R_k$. We give here a detailed derivation of how to obtain them. While this should not be seen as a mathematical proof, it is reasonable to think that it can be made rigorous along the lines of \cite{TW}.

\paragraph{Linear and non-linear relations for $\Phi_r$.}
The TBA-like integral equation system can be found from a set of relations that are satisfied by the previously defined $\Phi_r$. We establish them in this section. From now on we will omit to write the $p$-dependence of functions, and we will use the following shift notation:
\be
	f^{\pm s} \equiv \re^{\pm \ri s  \hbar \partial_p}f (p)=f(p \pm \ri s \hbar).
\ee
As a consequence of their definition, the $\phi$-functions in \eqref{phijdef} obey the following difference equation
\be
	\phi_j^{+1}-\phi_j^{-2}=\omega^{-1/2}V \phi_{j-1},
\ee
which is the inversion of (\ref{phijdef}).
On the $\Phi_{r}$ functions, this translates to
\be
\ba
	\label{Phidifference}
	\Phi_r^{+1}-\Phi_r^{-2} &=\kappa \omega^{-1/2}V \Phi_{r-1}, \\
	\bar \Phi_r^{-1}-\bar \Phi_r^{+2} &=\kappa \omega^{1/2}V \bar \Phi_{r-1}.
\ea
\ee
These are the linear relations satisfied by the $\Phi_r$. Another important set of relations that these functions satisfy is what we will call the ``quantum Wronskian relations''. This is a pair of non-linear equations which is given by 
\be
\ba
	\label{qw}
	\sum_{r=0}^{2} \omega^r \Phi_r^{+\frac{1}{2}} \bar \Phi_{3-r}^{-\frac{1}{2}}=1, \qquad \qquad
	\sum_{r=0}^{2} \omega^r \Phi_r^{-1} \bar \Phi_{3-r}^{+1}=1,
\ea
\ee
(again, the indices are defined modulo $3$). A motivation of their validity is given in appendix \ref{appQW}, let us just mention here that their origin lies in (\ref{zerocomb}). These relations are the equivalents of the Proposition in \cite{TW}, and might in principle be proven along similar lines.
Although we do not have a rigorous proof,
we have checked for 
several values of $\hbar$
that the quantum Wronskian relations \eqref{qw} are indeed satisfied in the small $\kappa$ 
expansion. 

\paragraph{Auxiliary functions $\eta_k$ and $\tilde \eta_k$.}
Let us define a set of auxiliary functions which will be useful when setting up the TBA system:
\be
\ba
	\label{etadef}
	\eta_k&=\omega^{\frac{3}{2}} \sum_{r=0}^{2} \omega^{r-\frac{k+1}{2}} \Phi_r^{+\frac{1}{2} } \bar \Phi_{k+1-r}^{-\frac{1}{2} }, \qquad \qquad
	\tilde \eta_k&=\omega^{\frac{3}{2}} \sum_{r=0}^{2} \omega^{r-\frac{k+1}{2}} \Phi_r^{-1 } \bar \Phi_{k+1-r}^{+1},
\ea
\ee
(as usual, indices are modulo $3$). These are non trivial definitions only for $k=0,1$, since $\eta_{2}=\tilde \eta_{2}=1$ as consequence of the quantum Wronskian relations. Using  (\ref{Phidifference}) and (\ref{nonvanishingRcomb}), we find
\be 
	\label{maineta}
	\eta_k^{+\frac{1}{2}}-\tilde \eta_k^{-1}=\eta_k^{-\frac{1}{2}}-\tilde \eta_k^{+1}=2\omega^{\frac{3}{2}} h \kappa \sin \left ( \frac{k+1}{3}\pi\right ) R_k.
\ee
By the same trick as the one we use in appendix \ref{appQW}, we can write the following relation between the two auxiliary functions:
\be
\ba
	\label{releta}
	 \eta_k=- \frac{ \sin \left (  \hbar \partial_p \right ) }{\sin \left ( \frac{  \hbar}{2} \partial_p \right )}\tilde \eta_k=-(\tilde \eta_k^{+\frac{1}{2}}+\tilde \eta_k^{-\frac{1}{2}}).
\ea
\ee
This allows us to rewrite (\ref{maineta}) as
\be 
	\label{finaleta}
	-\frac{\sin \left ( \frac{3}{2}\hbar \partial_p \right )}{\sin \left ( \frac{1}{2}\hbar \partial_p \right )} \tilde \eta_k =2\omega^{\frac{3}{2}} h \kappa \sin \left ( \frac{k+1}{3}\pi\right ) R_k.
\ee

\paragraph{TBA equations in difference form.} 
Let us build up the TBA equations using the ingredients that we previously defined.
We start by introducing the following ``building block'' \mbox{functions:}
\be
\ba
	\Psi=\sum_{r=0}^{2}\omega^{-r} \Phi_r, \qquad \qquad \bar  \Psi=\sum_{r=0}^{2} \omega^{r} \bar  \Phi_r.
\ea
\ee
Constructing the TBA system in difference form basically amounts to appropriately assembling and shifting these building blocks $\Psi$, and giving names to some of their combinations. Let us see how this works. The first important function appearing in the TBA system is
\be
	\mathcal R =\Psi \bar \Psi .
\ee
This can be expressed as
\be
\ba
	\label{Rmanip}
	\mathcal R &=\Psi \bar \Psi 
			=\sum_{r,s=0}^{2} \omega^{s-r} \Phi_r \bar \Phi_s =\sum_{k,\ell=0}^{2} \omega^{k-2 \ell} \Phi_\ell \bar \Phi_{k-\ell} 
			=\sum_{k=0}^2 \omega^{k+\frac{k}{2}} \sum_{\ell=0}^{2} \omega^{\ell-\frac{k}{2}} \Phi_\ell \bar \Phi_{k-\ell} \\
			&=\sum_{k=0}^1 \omega^{\frac{3}{2}k}\frac{2 h}{V} \sin \left (\pi \frac{k+1}{3} \right )R_k=\frac{\sqrt{3}}{V}(R_0-R_1) .
\ea
\ee
In the second line we used that since $\omega^3=1$, we have $\omega^{-2 \ell}=\omega^{\ell}$ for integer $\ell$, and in the third line, we used relations (\ref{zerocomb}) and (\ref{nonvanishingRcomb}). Now we use (\ref{finaleta}) to obtain
\be
\ba
	\kappa V \mathcal R &=-\frac{\sin \left( \frac{3}{2} \hbar \partial_p \right )}{ \sin \left( \frac{1}{2} \hbar \partial_p \right ) } \sum_{k=0}^{1}  			\omega^{\frac{3}{2}(k-1)} \tilde \eta_k
		=\left ( \re^{-\ri \hbar \partial_p} +1 + \re^{\ri \hbar \partial_p} \right ) \sum_{k=0}^{1}  \omega^{\frac{3}{2}k} \tilde \eta_k.
\ea
\ee
We define now the main auxiliary function $\mathcal H$ as 
\be
	\label{Hdef}
	\mathcal H =   \sum_{k=0}^{1} \omega^{\frac{3}{2}k}  \tilde \eta_k,
\ee
and thus arrive at the first TBA-like equation (in difference form) :
\be
\ba
	\kappa V \mathcal R &=  \mathcal H^{-1}+\mathcal H+\mathcal H^{+1}.
\ea
\ee
To arrive at the second TBA equation, we compute the following products using  the quantum Wronskian relations, (\ref{etadef}) and (\ref{releta}):
\be
\ba
	\Psi^{+\frac{1}{2}} \bar \Psi^{-\frac{1}{2}} &= \sum_{k,\ell}^2 \omega^{k-2\ell}  \Phi_{\ell}^{+\frac{1}{2}} \bar  \Phi_{k-\ell}^{-\frac{1}{2}} = 1+\sum_{k=0}^1 \omega^{\frac{3}{2}k} \eta_k = 1-\sum_{k=0}^1 \omega^{\frac{3}{2}k} \eta_k ^{+\frac{1}{2}}-\sum_{k=0}^1 \omega^{\frac{3}{2}k} \eta_k ^{-\frac{1}{2}} \\
		&=1-\mathcal H^{-\frac{1}{2}}-\mathcal H ^{+\frac{1}{2}},\\
	\Psi^{-1} \bar \Psi^{+1} &= \sum_{k,\ell}^2 \omega^{k-2\ell}  \Phi_{\ell}^{-1} \bar  \Phi_{k-\ell}^{+1} = 1+\sum_{k=0}^1 \omega^{\frac{3}{2}k} \eta_k \\
		&=1+\mathcal H.
\ea
\ee
It is now just a matter of shifting these relations to get the second TBA equation:
\be
\ba
	{\mathcal R}^{-1} {\mathcal R} {\mathcal R}^{+1} &=\Psi^{-1} \bar \Psi^{-1} \Psi \bar \Psi \Psi^{+1} \bar \Psi^{+1}=(\Psi^{-1} \Psi^{+1})(\Psi^{-\frac{1}{2}} \bar \Psi^{-\frac{1}{2}})^{-\frac{1}{2}} (\Psi^{+\frac{1}{2}} \bar \Psi^{-\frac{1}{2}})^{+\frac{1}{2}} \\
	&=(1+\mathcal H)(1-\mathcal H^{-1}-\mathcal H)(1-\mathcal H-\mathcal H^{+1}).
\ea
\ee
We arrived at a system of two difference equations satisfied by functions $\mathcal R$ and $\mathcal H$, which are  in principle determined by these equations. Once they are solved, some information on the traces can be extracted from $\mathcal R$ ($\mathcal H$ is just an auxiliary function). But this is not enough to retrieve the whole diagonal part of the resolvent function, since $\mathcal R$ does not contain $R_2$.  To obtain that, we need to introduce yet another function :
\be
\ba
	\mathcal W= \{\Psi,\bar \Psi\}=\ri (\Psi' \bar \Psi-\Psi \bar \Psi'),
\ea
\ee
where the derivative is taken with respect to the variable $p$. From this it follows that
\be
\ba
	\frac{\mathcal W}{\mathcal R}=\ri \frac{\partial}{\partial p} \log \frac{\Psi}{\bar \Psi},
\ea
\ee
which leads us to the last equation of the TBA-like system :
\be
	\label{3rdtba}
\ba
	\left ( \frac{\mathcal W}{\mathcal R} \right )^{+1}-\left ( \frac{\mathcal W}{\mathcal R} \right )^{-1} & =\ri \frac{\partial}{\partial p} \log \frac{\Psi^{+1} \bar \Psi^{-1}}{\bar \Psi^{+1} \Psi^{-1}} = \ri \frac{\partial}{\partial p} \log \frac{(\Psi^{+\frac{1}{2}} \bar \Psi^{-\frac{1}{2}})^{-\frac{1}{2} } (\Psi^{+\frac{1}{2}} \bar \Psi^{-\frac{1}{2}})^{+\frac{1}{2} }}{(\Psi^{-1} \bar \Psi^{+1} )(\Psi \bar \Psi)}  \\ 
	&= \ri \frac{\partial}{\partial p} \log \frac{(1-\mathcal H^{-1}-\mathcal H)(1-\mathcal H-\mathcal H^{+1})}{\mathcal R (1+\mathcal H)} \\
	&=\ri \frac{\partial}{\partial p} \log \frac{\mathcal R^{+1} \mathcal R^{-1}}{\mathcal (1+\mathcal H)^2} .
\ea
\ee
Let us summarize the TBA system, written here in the very useful difference form:
\be
	\label{p2TBA}
	\begin{cases} 
		\kappa V \mathcal R =  \mathcal H^{-1}+\mathcal H+\mathcal H^{+1} \\ 
		{\mathcal R}^{-1} {\mathcal R} {\mathcal R}^{+1} =(1+\mathcal H)(1-\mathcal H^{-1}-\mathcal H)(1-\mathcal H-\mathcal H^{+1}) \\ 
		\left ( \frac{\mathcal W}{\mathcal R} \right )^{+1}-\left ( \frac{\mathcal W}{\mathcal R} \right )^{-1} = \ri \frac{\partial}{\partial p} \log \frac{\mathcal R^{+1} \mathcal R^{-1}}{\mathcal (1+\mathcal H)^2} .
	 \end{cases} 
\ee
Once this system is solved (which is of course far from trivial!), it is straightforward to extract the diagonal $R_k$ functions (\ref{Rkdef}) from the solution $(\mathcal R, \mathcal H, \mathcal W)$:
\be
\ba
	\label{RkfromTBA}
	R_k&=\frac{(-1)^k V}{18 \hbar \sin \left ( \frac{k+1}{3}\pi \right )} \sum_{r=0}^{2} \omega^{-r k} \mathcal R(\omega^r \kappa), \qquad \qquad k=0,1 \\
	R_{2}&=\frac{V}{12 \pi } \sum_{r=0}^{2} \omega^{r} \mathcal W(\omega^r \kappa),
\ea
\ee
where we remind that $\omega=\re^{\frac{2\pi \ri}{3} }$.

\paragraph{TBA equations in integral form.} 
To write the integral TBA equations, we invert the difference equations given in (\ref{p2TBA}). Let us define
\be
\ba
	\epsilon=-\log \mathcal R.
\ea
\ee
Then, we can write the system in the following way:
\be
\ba
	\label{p2TBAint}
		\mathcal H(p) &=  \frac{\kappa}{\sqrt{3}\hbar} \int_{-\infty}^{\infty} \rd p' \,  \frac{\sinh \left (\frac{\pi}{3}\frac{p-p'}{\hbar} \right )}{\sinh \left (\pi \frac{p-p'}{\hbar}\right )}  V(p') \re^{-\epsilon(p')},  \\ 
		\epsilon (p) &=  -\frac{1}{\sqrt{3}\hbar} \int_{-\infty}^{\infty} \rd p' \,  \frac{\sinh \left (\frac{\pi}{3}\frac{p-p'}{\hbar} \right )}{\sinh \left (\pi \frac{p-p'}{\hbar}\right )}  \log \Big [ \left (1+\mathcal H(p') \right )    \left (1-\mathcal H(p'-\ri \hbar)-\mathcal H(p') \right )\\[-0.5cm]
		& \qquad \qquad \qquad \qquad \qquad \qquad \qquad  \qquad \qquad  \qquad \qquad  
		    \left (1-\mathcal H(p')-\mathcal H(p'+\ri \hbar) \right ) \Big ],  \\ 
		    \mathcal W(p) &= \frac{\pi}{8 \hbar^2} \re^{-\epsilon(p)} \int_{-\infty}^{\infty} \rd p' \,  \frac{1}{\cosh^2 \left ( \frac{\pi}{2} \frac{p-p'}{\hbar}\right )}  \log \left [ \frac{\re^{-\epsilon(p'+\ri \hbar)-\epsilon(p'-\ri \hbar)}}{(1+\mathcal H(p'))^2} \right].
\ea
\ee
This is the main result of this section. The structure of this system is quite similar to the $C=0$ case, but involves different integral kernels, and more complicated combinations of the auxiliary function $\mathcal H$. To retrieve the form that is given in the Introduction, we just need to rescale (and rename) some of the functions, and hide the $\hbar$ dependence in the potential $V$ which is arbitrary anyway. We also use the elementary fact that 
\be
	\frac{\sinh \frac{x}{2}}{\sinh \frac{3x}{2}}=\frac{1}{1+2\cosh x}.
\ee
 

\paragraph{A note concerning the potential.} As we explained, the TBA system given in (\ref{p2TBA}) is valid for any well behaved real potential. Let us now return for the particular case of local $\mathbb P^2$. In that case, we must consider a family of potentials which depend on $\hbar$ as
 \be
   	V({ p})=W({ p}) \overline{ W({ p})}=\re^{\frac{p}{3}} \frac{\Phi_b(\frac{p+\ri \hbar}{2 \pi b})}{\Phi_b(\frac{p-\ri \hbar}{2 \pi b})}, \qquad \qquad  b^2=\frac{3 \hbar}{2\pi}.
\ee
To obtain the above expression, we used the unitary relation of the Faddeev quantum dilogarithm: $1/\Phi_b(z)=\Phi^*_b(z)\equiv \overline{\Phi_b(\bar z)}$. For some fixed values of $\hbar$, notably $\pi$ times a rational number, this expression can be simplified to more common functions (see some examples in section \ref{sec:exact-Z}). But the general $\hbar$ dependence of the potential is through this rather complicated function $\Phi_b$. It is therefore interesting to notice that, in a way, the structure of the TBA system partially ``untangles'' this complicated $\hbar$-dependence, at least as far as the auxiliary function $\mathcal H$ is concerned. Indeed, if we express the first two equations of (\ref{p2TBA}) solely in terms of $\mathcal H$, we obtain
 \be
 	\label{Hfunctionaleq}
	\frac{1}{\kappa^3 V^{-1}  V V^{+1} }  = \frac{(1+\mathcal H)(1-\mathcal H^{-1}-\mathcal H)(1-\mathcal H-\mathcal H^{+1})}{(\mathcal H^{-2}+\mathcal H^{-1}+\mathcal H)(\mathcal H^{-1}+\mathcal H+\mathcal H^{+1})(\mathcal H+\mathcal H^{+1}+\mathcal H^{+2})}.	
\ee
The particular combination of $V$ appearing above is given by
 \be
	V^{-1}  V V^{+1}   = \re^{p}\frac{\Phi_b(\frac{p}{2\pi b}+\frac{\ri b}{3}) \Phi_b(\frac{p}{2\pi b}+\frac{2\ri b}{3})}{\Phi_b(\frac{p}{2\pi b}-\frac{2\ri b}{3})\Phi_b(\frac{p}{2\pi b}-\frac{\ri b}{3})}.
\ee
But this can be seen to reduce to a simple function for any value of $\hbar$ by using the functional relation of the Faddeev quantum dilogarithm
\be
	\frac{\Phi_b(\frac{x}{2\pi b}+\frac{\ri b}{2})}{\Phi_b(\frac{x}{2\pi b}-\frac{\ri b}{2})}=\frac{1}{\re^{x}+1}.
\ee
Indeed, we find
\be
	\label{3V}
	\frac{1}{V(p-\ri \hbar)  V(p) V(p+\ri \hbar)}=2 \cosh(p)+2\cos(\hbar/2).
\ee
So it seems that, even if the family of potentials relevant for the local $\mathbb P^2$ case is complicated, it has a sort of ``compatibility'' feature with respect to the TBA system. However, even considering this simplification, finding the solution(s) of (\ref{Hfunctionaleq}) with (\ref{3V}) is still a complicated task.

\subsection{Semiclassical expansion of the solution}

Finding an explicit solution for the TBA equations (\ref{p2TBAint}) is challenging. Given a potential $V$, it can however be done order by order in $\kappa$ (exactly in $\hbar$), thus yielding the first few exact traces of the operator $\rho$ in a recursive manner. This method was used in \cite{PY} for the $C=0$ case, and can be readily adapted to our system.
Some of these results for our TBA are given in section \ref{sec:exact-Z}.
In this section, we will do something different: we will expand the solution in small $\hbar$ and exactly in $\kappa$, in the spirit of \cite{CM}. This yields the semiclassical expansion of all the traces of the operator $\rho$, which can be compared for example to the WKB calculations of \cite{HW}. Also, this allows us to compare with the refined topological data of local $\mathbb P^2$, as will be explained later.

In order to expand the TBA system at small $\hbar$, we will firstly need the semiclassical expansion of the potential $V$ appearing in (\ref{P2kernel}). This can easily be done by using the semiclassical expansion of the Faddeev quantum dilogarithm \cite{AK}. The result can be written as
\be
\ba
	V(p) & ={\rm exp} \left (  \frac{p}{3} +\frac{2}{3\hbar} \sum_{n \geq 0}(-9 \hbar^2)^n \frac{B_{2n}(\frac{1}{2})}{(2n)!} {\rm Im \, Li}_{2-2n}(-\re^{p+\ri \hbar}) \right ) \\
		&=\frac{1}{(2 \cosh \frac{p}{2})^{\frac{2}{3}}}-\frac{5}{36} \frac{1}{(2 \cosh \frac{p}{2})^{\frac{8}{3}}} \hbar^2 + \frac{242-153 \cosh^2 \frac{p}{2}}{1296} \frac{1}{(2 \cosh \frac{p}{2})^{\frac{14}{3}}} \hbar^4+ O(\hbar^6).
\ea
\ee
We now have all the ingredients to perform the systematic semiclassical expansion of the TBA equation. We expand the solutions as follows:
\be
\ba
	\mathcal H(p)= \sum_{\ell \geq 0} h_\ell(p) \hbar^{2\ell},  \qquad 
	 \mathcal R(p)= \sum_{\ell \geq 0} r_\ell(p) \hbar^{2\ell},  \qquad
	 \frac{\mathcal W(p)}{\mathcal R(p)}=\frac{1}{\hbar }\sum_{\ell \geq 0} t_\ell(p) \hbar^{2\ell}.
\ea
\ee

\paragraph{Leading order in the semiclassical expansion.} At leading order in $\hbar$, all the shifted functions in (\ref{p2TBA}) can be replaced by their unshifted version since the shift is proportional to $\hbar$. Also, the potential $V$ is to be replaced by its leading-$\hbar$ expression. In this limit, the semiclassical TBA becomes algebraic :
\be
	\label{p2TBAsemiclass}
	\begin{cases} 
		3  h_0 =\frac{\kappa}{(2 \cosh \frac{p}{2})^{2/3 }}  r_0  \\ 
		{ r_0}^{3}  =(1+ h_0)(1-2  h_0)^2 \\ 
		t_0= \log \frac{r_0}{1+h_0}.
	 \end{cases} 
\ee
The first two lines yield a cubic equation for $h_0$ :
\be
	4 \left (1-a^{-3} \right ) h_0^3-3 h_0 +1=0,
\ee
where
\be
	a= \frac{\kappa}{3 \cosh^{2/3} \frac{p}{2}}.
\ee
The solution we are interested in is given by
\be 
\ba
	h_0=\frac{a(1+\sqrt{1-a^3})^{2/3}-a^2}{2\sqrt{1-a^3} (1+\sqrt{1-a^3})^{1/3}} ,   \qquad
  	 r_0=\frac{2^{\frac{2}{3}}}{a}h_0, \qquad
	t_0 &= \log \frac{r_0}{1+h_0}.
\ea
\ee
Using (\ref{RkfromTBA}), we reconstruct the diagonal resolvents in the leading semiclassical approximation:
\be
\ba
	R_0 &= \frac{a \left ( \sqrt{1-a^3}+1 \right)^{\frac{1}{3}} }{2 \sqrt{3} \kappa  \sqrt{1- a^3}  } \frac{1}{\hbar}+ \mathcal O(\hbar)=\frac{1}{  2^{2/3}  \sqrt{3} \kappa} a \, _2F_1\left(\frac{1}{3},\frac{5}{6};\frac{2}{3};a^3\right) \frac{1}{\hbar}+ O(\hbar), \\[0.2cm]
	R_1 &= \frac{a^2 \left ( \sqrt{1-a^3}+1 \right)^{-\frac{1}{3}} }{2 \sqrt{3} \kappa  \sqrt{1- a^3}  } \frac{1}{\hbar}+ \mathcal O(\hbar) = \frac{1}{  2^{4/3}  \sqrt{3} \kappa}  a^2 \, _2F_1\left(\frac{2}{3},\frac{7}{6};\frac{4}{3};a^3\right)  \frac{1}{\hbar}+  O(\hbar) , \\[0.3cm]
	R_2 &=\frac{2^{1/3}a}{8\pi \kappa}\left ( r_0(a)t_0(a)+\re^{\frac{2\pi \ri}{3}}  r_0(\re^{\frac{2\pi \ri}{3}} a)t_0(\re^{\frac{2\pi \ri}{3}} a)+\re^{\frac{4\pi \ri}{3}} r_0(\re^{\frac{4\pi \ri}{3}}  a)t_0(\re^{\frac{4\pi \ri}{3}}  a)\right ) \frac{1}{\hbar}+  O(\hbar) \\
	& \qquad \qquad \qquad \qquad  \qquad \qquad  \qquad  \,\,
	=\frac{9}{16 \pi  \kappa }  a^3 \, _3F_2\left(1,1,\frac{3}{2};\frac{4}{3},\frac{5}{3};a^3\right) \frac{1}{\hbar}+  O(\hbar).
\ea
\ee
According to (\ref{Rkrelations}), summing all the $R_k$ and integrating over $p$ (taking the trace in the $p$ representation) yields the generating function of the semiclassical traces:
\be
\ba
	\hbar \,\, \tr \, {\rm R} &=\int_{-\infty}^{\infty} \rd p \, (R_0+R_1+R_2)=2 \int_{0}^{\kappa/3} \frac{3 \, \rd a}{a \sqrt{1-(3a/\kappa)^3}}(R_0+R_1+R_2) \\
	&=
	\frac{\Gamma \left(\frac{1}{3}\right)^3 \,
   _2F_1\left(\frac{1}{3},\frac{1}{3};\frac{2}{3};\frac{\kappa ^3}{27}\right)}{6 \pi }
	+\frac{\kappa 
   \Gamma \left(\frac{2}{3}\right)^3 \, _2F_1\left(\frac{2}{3},\frac{2}{3};\frac{4}{3};\frac{\kappa
   ^3}{27}\right)}{6 \pi }
   +\frac{\kappa ^2 \, _3F_2\left(1,1,1;\frac{4}{3},\frac{5}{3};\frac{\kappa ^3}{27}\right)}{12 \pi }+ O(\hbar^2).
\ea
\ee
We can also obtain the semiclassical grand-potential through (\ref{RtoJ}) :
\be
\ba
	\label{leadingJ}
	\hbar \,\, \mathcal J(\kappa) &= \frac{\kappa  \Gamma \left(\frac{1}{3}\right)^3 \,
   _3F_2\left(\frac{1}{3},\frac{1}{3},\frac{1}{3};\frac{2}{3},\frac{4}{3};-\frac{\kappa ^3}{27}\right)}{6 \pi
   }
   -\frac{\kappa ^2 \Gamma \left(\frac{2}{3}\right)^3 \,
   _3F_2\left(\frac{2}{3},\frac{2}{3},\frac{2}{3};\frac{4}{3},\frac{5}{3};-\frac{\kappa ^3}{27}\right)}{12 \pi }\\
   & \qquad \qquad \qquad \qquad \qquad \qquad \qquad \qquad   
   +\frac{\kappa ^3 \, _4F_3\left(1,1,1,1;\frac{4}{3},\frac{5}{3},2;-\frac{\kappa ^3}{27}\right)}{36 \pi
   }+ O(\hbar^2),
  \ea
\ee
which reproduces the result in \cite{GHM}.

\paragraph{Subleading orders.} The subleading orders in $\hbar$ of the TBA equations yield recursive equations for the $h_\ell$ in terms of lower ones and their derivatives. For example, $h_1$ is given by
\be
\ba
	h_1 &= \frac{-\frac{27 a^3 h_0^3}{4 \kappa ^3}+h_0'' \left(2 \left(a^3-4\right) h_0^2+a^3
   (h_0-1)\right)+h_0'^2 \left(\left(4-a^3\right) h_0-a^3\right)}{3 \left(4
   \left(a^3-1\right) h_0^2-a^3\right)},
\ea
\ee
where the primes are derivatives with respect to $p$. With this method, the TBA can be solved to high order in $\hbar$. Here we present the $\hbar$-corrections for the traces:
\be
\ba
	\label{tracecorr}
	\tr \rho^n&=\frac{\Gamma (\frac{n}{3})^3}{6\pi \hbar \Gamma(n)} \Big  (1-\frac{n^2}{72} \hbar^2 +\frac{n^3(7n+16)}{51840} \hbar^4 \mathcal -\frac{n^3(93 n^3+656 n^2 +1536 n+1152)}{78382080} \hbar^6+ \\
	& \qquad 
	+\frac{n^3(1143 n^5+16480 n^4+97536 n^3 +292608 n^2+442368 n +269568)}{112 870 195 200} \hbar^8
	 +O(\hbar^{10}) \Big ).
	 \\[-0.4cm]
	 \,
\ea
\ee
If we express the leading order of $\mathcal J$ in (\ref{leadingJ}) as $\mathcal J_0(\mu)$ with $\kappa=\re^{\mu}$, it is easy to see that $\hbar$-corrections can be encoded in differentials applied on the leading semiclassical solution:
\be
\ba
	\label{Jsolhbar}
	\hbar \, \mathcal J(\mu)=\left ( 1+ \sum_{\ell \geq 1} \hbar ^{2 \ell} D_{\ell}(\partial_\mu) \right) \hbar \, \mathcal J_0(\mu),
\ea
\ee
where $D_n$ are the polynomials appearing in (\ref{tracecorr}):
\be
\ba
	D_{1}(\partial_\mu)=-\frac{\partial_\mu^2}{72}, \qquad  D_{2}(\partial_\mu)=\frac{\partial_\mu^3(7\partial_\mu+16)}{51840},
	\qquad  D_{3}(\partial_\mu)= -\frac{\partial_\mu^3(93 \partial_\mu^3+656 \partial_\mu^2 +1536 \partial_\mu+1152)}{78382080}, 
\label{Dn}
\ea
\ee
and so on. These differentials  are exactly those given in \cite{Hat} (for $m=n=1$).

In \cite{Hua}, 
the WKB expansion of quantum periods for local $\mathbb{P}^2$
are expressed in terms of certain differential operators.
We can see the relation between the differential operators in \cite{Hua}
and those in \eqref{Dn}, as follows.
Let us consider the WKB expansion of the quantum periods of local $\mathbb{P}^2$
\begin{align}
 t=\sum_{n=0}^\infty t_{2n} \hbar^{2n}.
\end{align}
The leading term $t_0$ is a classical period satisfying the Picard-Fuchs equation
\begin{align}
 \Theta_z^3 t_0+3z(3\Theta_z+2)(3\Theta_z+1)\Theta_z t_0=0,
\end{align}
where $\Theta_z=z\partial_z$ and $z=e^{-3\mu}$.
The correction terms were found to be \cite{Hua}\footnote{Our definition of $\hbar$ is different from \cite{Hua} by a factor
of $\ri$, hence there is an extra sign $(-1)^n$ for $t_{2n}$ compared to \cite{Hua}.}
\begin{equation}
\begin{aligned}
 t_2&=-\frac{\Theta_z^2t_0}{8},\\
t_4&=\frac{2z(999z-5)\Theta_zt_0+3z(2619z-29)\Theta_z^2t_0}{640\Delta^2},
\label{t-Huang}
\end{aligned} 
\end{equation}
with $\Delta=1+27z$.
Using the relations following from the Picard-Fuchs equation
\begin{equation}
\begin{aligned}
 \Theta_z^3 t_0&=-\frac{3z}{\Delta}(9\Theta_z^2 t_0+2\Theta_z t_0),\\
\Theta_z^4 t_0&=-\frac{3z}{\Delta^2}\Big[(189z-11)\Theta_z^2 t_0+2(27z-1)\Theta_z t_0\Big],
\end{aligned} 
\end{equation}
one can show that \eqref{t-Huang} can be rewritten as
\begin{align}
 t_2=-\frac{\partial_\mu^2}{72}t_0,\qquad
t_4=\frac{\partial_\mu^3(7\partial_\mu+16)}{51840}t_0,
\end{align}
which agree with the differential operators in \eqref{Dn}.

\paragraph{BPS invariants and TBA.} The small-$\hbar$ expansion of the grand
potential is known to encode some refined topological data of the CY geometry. Let us check this fact in our solution. In the language of \cite{HMMO,GHM}, 
\be
\ba
	\label{Jfromtop}
	\mathcal J(\mu)=J^{(\rm p)}(\mu_{\rm eff})+J^{(\rm WKB)}(\mu_{\rm eff})+\text{non. pert.} ,
\ea
\ee
where ``non. pert.'' is the non-perturbative part in $\hbar$ which is invisible in the small-$\hbar$ expansion, corresponding to the worldsheet instanton
corrections in the standard topological string on local $\mathbb P^2$. 
It will not matter in our present study. 
On the other hand, $J^{(\rm WKB)}$
corresponds to the ``membrane instantons'', which is related to the
refined topological string in the Nekrasov-Shatashvili limit \cite{HMMO}. 
The explicit forms of $J^{(\rm p)}$ and $J^{(\rm WKB)}$ are given by
\be
\ba
	\label{Jmuparts}
	J^{(\rm p)}(\mu) &=\frac{C(\hbar)}{3} \mu^3+B(\hbar)\mu +A(\hbar), \\
	J^{(\rm WKB)}(\mu) &=\sum_{\ell \geq 1} \re^{-r \ell \mu} \left ( \mu +\frac{1}{r \ell}-\frac{\hbar \partial_\hbar}{r \ell} \right ) \tilde b_{\ell}(\hbar).
\ea
\ee
Here $r=3$, and the coefficients $C(\hbar),B(\hbar)$ and $A(\hbar)$ 
in $J^{(\rm p)}(\mu)$ are given by \cite{GHM}
\begin{equation}
   C(\hbar)=\frac{9}{4\pi\hbar},\qquad
B(\hbar)=\frac{\pi}{2\hbar}-\frac{\hbar}{16\pi},\qquad
A(\hbar)=\frac{3}{4}A_c\Bigl(\frac{\hbar}{\pi}\Bigr)-\frac{1}{4}A_c\Bigl(\frac{3\hbar}{\pi}\Bigr),
\end{equation}
where $A_c(k)$ denotes the constant term in the ABJM theory.
In \eqref{Jfromtop}, $\mu_{\rm eff}$ is related to $\mu$ through the quantum mirror map $\Pi_A(z;\hbar)$ \cite{HMMO}, which is the quantum version of the mirror map relating the K\"ahler parameter to the complex structure deformation \cite{ACDKV}. The  $\tilde b_{\ell}(\hbar)$ encodes some combinations of the BPS invariants $N_{j_L,j_R}^d $, (which can be seen as some kind of refined enumerative invariants):
\be
\ba
	\re^{-r \mu_{\rm eff}} &=-{\rm exp}\left ( \Pi_A(-\re^{-r \mu}; \hbar) \right) , \\
	\tilde b_{\ell}(\hbar) &= -\frac{r \ell}{4 \pi} \sum_{j_L,j_R}  \sum_{\substack{d,w \\ dw=\ell}} (-1)^{2j_L +2 j_R}N_{j_L,j_R}^d \frac{ \sin \left (\frac{w\hbar}{2}(2 j_L+1 \right ) \sin \left (\frac{w\hbar}{2}(2 j_R+1) \right )}{w^2 \sin^3 \left (\frac{w\hbar}{2} \right )}.
\ea
\ee
 The $N_{j_l,j_R}^d$ for local $\mathbb P^2$ can be found for example in Table 7 of \cite{HW}. Expressions (\ref{Jsolhbar}) and (\ref{Jfromtop}) can be compared if we do a large $\mu$ expansion of our solution (\ref{Jsolhbar}). This can be done by using the Mellin-Barnes integral representation of our solution, as in \cite{Hat}. We start with the leading term of the grand potential, which is given by
\be
	\mathcal J_0(\mu) =\sum_{n \geq 1} \frac{\Gamma (\frac{n}{3})^3}{6\pi \Gamma(n)}\re^{n\mu}=-\int_{-\ri \infty +\epsilon}^{\ri \infty+\epsilon} \frac{\rd s}{2 \pi \ri} \Gamma(-s)\frac{\Gamma (\frac{s}{3})^3}{6\pi} \re^{s \mu}.
\ee
Picking up the residues at $s=-3\ell \,\, (\ell \leq 0)$, we can easily find the large $\mu$
 expansion of $\mathcal J_0(\mu)$:
\be
	\mathcal J_0(\mu) =\frac{3 \mu^3}{4 \pi}+\frac{\pi \mu}{2}+\frac{4 \zeta(3)}{3 \pi}+\sum_{\ell=1}^{\infty}(a_\ell^{(0)}\mu^2+b^{(0)}_{\ell} \mu+c_{\ell}^{(0)})\re^{-3 \ell \mu},
\ee
where
\be
\ba
	a^{(0)}_\ell&=\frac{9(-1)^{\ell-1}}{4\pi} \frac{(3\ell-1)!}{(\ell !)^3} ,\\
	b^{(0)}_\ell&=\frac{9(-1)^{\ell-1}}{2\pi^2} \frac{(3\ell-1)!}{(\ell !)^3}  \left [ \psi(3\ell)+\psi(\ell+1) \right ],\\
	c^{(0)}_\ell&=\frac{(-1)^{\ell-1}}{4\pi} \frac{(3\ell-1)!}{(\ell !)^3}  \left [\pi^2+9 \left ( \psi(3\ell)+\psi(\ell+1) \right)^2+9 \psi^{(1)}(3\ell)-3 \psi^{(1)}(\ell+1) \right ].
\ea	
\ee
In the above, $\psi(z)$ is the digamma function and $\psi^{(n)}(z)$ is its $n^{\rm th}$ derivative. From (\ref{Jsolhbar}), we have that the $\ell$-instanton coefficients at order $n$ are given by
\be
\ba
	a^{(n)}_\ell&=a^{(0)}_\ell D_n(-3\ell) ,\\
	b^{(n)}_\ell&=b^{(0)}_\ell  D_n(-3\ell)+2 a^{(0)}_\ell  D_n'(-3 \ell) ,\\
	c^{(n)}_\ell&=c^{(0)}_\ell  D_n(-3\ell)+b^{(0)}_\ell  D_n'(-3 \ell)+a^{(0)}_\ell  D_n''(-3 \ell).
\ea	
\ee
We have checked the agreement between (\ref{Jsolhbar}) and 
\eqref{Jmuparts} up to 6-instanton and $\mathcal O(\hbar^{40})$.
This gives strong evidence for the conjecture in \cite{HMMO,GHM} for the local $\mathbb{P}^2$
case, that the membrane instanton corrections are indeed given by the refined topological string
in the Nekrasov-Shatashvili limit.
To study the worldsheet instanton corrections, we need to go beyond the small $\hbar$ expansion,
which we will consider in the next section.

\section{Exact traces for local $\mathbb P^2$}
\label{sec:exact-Z}
For a fixed value of $\hbar$ chosen to be a rational number times $\pi$, the TBA system (\ref{p2TBAint}) can be solved order by order in $\kappa$ exactly as done in \cite{PY} (the computation consists of recursively computing integrals by taking residues). For $\hbar = 2\pi$, the Faddeev quantum dilogartihm simplifies, so that the $V$ term in (\ref{P2kernel}) can be written as
 \be
 	V(p)=\frac{1}{1+2\cosh\frac{p}{3}}=\frac{\sinh \frac{p}{6}}{\sinh \frac{p}{2} }.
 \ee
The first 10 traces are
\be
\ba
	\nonumber
	\tr \rho \, \, \, &=\frac{1}{9} ,\qquad \qquad \qquad \qquad \qquad \qquad \qquad \quad \,\,
	\tr \rho^2 =\frac{1}{27}-\frac{1}{6 \sqrt{3} \pi } ,\\
	\tr \rho^3 &= \frac{1}{81}-\frac{1}{24 \sqrt{3} \pi }-\frac{1}{24 \pi ^2}, \qquad \qquad \qquad \quad
	\tr \rho^4 = -\frac{1}{729}+\frac{1}{72 \pi ^2} ,\\
	\tr \rho^5 &=-\frac{2}{2187}+\frac{7}{3888 \sqrt{3} \pi }+\frac{5}{864 \pi ^2}, \qquad \quad \,\,\,\,
	\tr \rho^6 = -\frac{1}{2187}+\frac{1}{720
   \sqrt{3} \pi }+\frac{1}{432 \pi ^2}-\frac{1}{576 \sqrt{3} \pi ^3}, \\
	\tr \rho^7 &= \frac{2}{59049}-\frac{49}{233280 \pi ^2}-\frac{7}{10368 \sqrt{3} \pi ^3}, \\
	\tr \rho^8 &= \frac{5}{177147}-\frac{71}{2755620 \sqrt{3} \pi }-\frac{17}{87480 \pi ^2}-\frac{1}{3888 \sqrt{3} \pi
   ^3}+\frac{1}{10368 \pi ^4}.
   \ea
   \ee
   \be
\ba
   	\tr \rho^9 &= \frac{1}{59049}-\frac{1}{20160 \sqrt{3}
   \pi }-\frac{43}{483840 \pi ^2}+\frac{7}{207360 \sqrt{3} \pi ^3}+\frac{1}{18432 \pi ^4}, \\
	\tr \rho^{10} &= -\frac{14}{14348907}+\frac{1763}{529079040 \pi ^2}+\frac{41}{1679616 \sqrt{3} \pi ^3}+\frac{5}{248832 \pi
   ^4}-\frac{1}{82944 \sqrt{3} \pi ^5}, \qquad \qquad \quad \quad \\
\ea
\ee
\normalsize
In the case where $\hbar=4\pi$, the $V$ term is also simplified. It is given by
 \be
 	V(p)=\frac{\cosh \frac{p}{3}+\frac{\sqrt{3}}{2}}{\cosh p}.
 \ee
 The first few traces are
\be
\ba
	\tr \rho \, \, \, &=\frac{1}{36} ,\\
	\tr \rho^2 &=\frac{1}{6 \sqrt{3} \pi }-\frac{13}{432} ,\\
	\tr \rho^3 &= -\frac{121}{10368}+\frac{1}{48 \pi ^2}+\frac{5}{96 \sqrt{3} \pi }, \\
	\tr \rho^4 &= \frac{113}{93312}+\frac{1}{144 \pi ^2}-\frac{1}{96 \sqrt{3} \pi } ,\\
	\tr \rho^5 &=-\frac{8033}{8957952}+\frac{25}{6912 \pi ^2}+\frac{359}{124416 \sqrt{3} \pi }.\\[0.1cm]
\ea
\ee
The first (highest) eigenvalue of this operator which we note $\re^{-\lambda_0}$  is given in \cite{GHM}. As an amusing check, let us use our five traces to approximate it by numerical methods (minimization, fixed point method, etc.). We find
\be
 	\re^{-\lambda_0}=3.777706(4),
 \ee
which is to be compared with the exact value
\be
 	\re^{-\lambda_0}=3.7777062582... \, .
 \ee
In the case where $\hbar=6\pi$, the $V$ term is given by
 \be
 	V(p)=\frac{\sinh \frac{p}{6}}{\sinh \frac{3p}{2} }\left ( \cosh p - \cos \frac{2 \pi}{9} \right ).
 \ee
 The very first traces are
\be
\ba
	\label{6pitraces}
	\tr \rho &= -\frac{2}{27}-\frac{\sin \left(\frac{\pi
   }{9}\right)}{9 \sqrt{3}}+\frac{1}{9} \cos
   \left(\frac{\pi }{9}\right), \\
   	\tr \rho^2 &=\frac{4}{81}-\frac{2}{81} \sin \left(\frac{\pi }{18}\right)-\frac{\sqrt{21+6 \sqrt{39} \cos \left(\frac{1}{3} \arctan
   \left(\frac{19}{53 \sqrt{3}}\right)\right)}}{54 \pi }.
  \ea
\ee
In the case $\hbar=\pi$, the $V$ term is given by
 \be
 	V(p)=\frac{1}{2\cosh\frac{p}{3}}.
 \ee
The first traces are
\be
\ba
 	\tr \rho &= \frac{1}{2\sqrt{3}}, \\
	\tr \rho^2 &= \frac{1}{36}, \\
	\tr \rho^3 &= -\frac{1}{16\sqrt{3}}+\frac{1}{8\pi}, \\
	\tr \rho^4 &=-\frac{13}{432}+\frac{1}{6\sqrt{3}\pi}.
\ea
 \ee
For  $\hbar=4\pi$ and $\hbar=6\pi$ we used equations (\ref{phijdef}) and (\ref{powerkernel}), which usually proved to be more effective than iterating up the integral TBA system.

\begin{figure}[htb]
\begin{center}
\includegraphics[width=8cm]{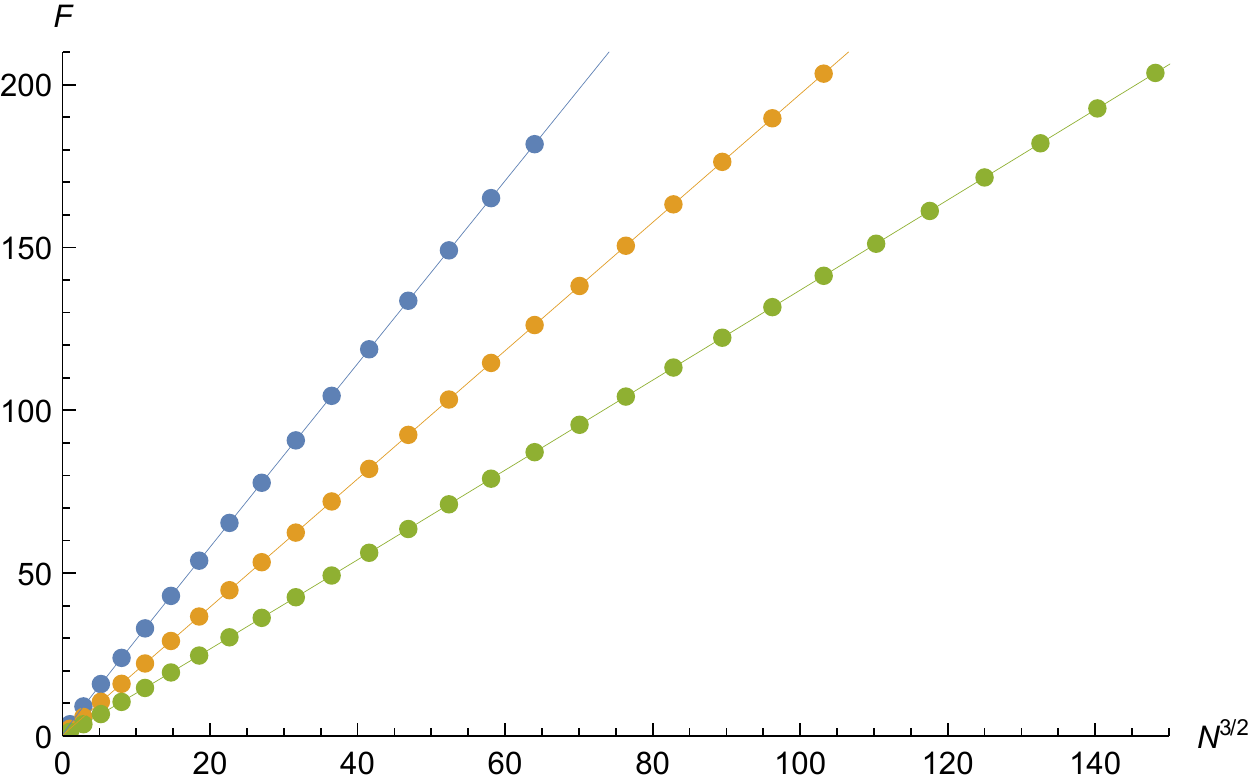}
\end{center}
  \caption{We show  the plot of free energy $F=-\log Z(N,\hbar)$ of local $\mathbb{P}^2$,
for $\hbar=4\pi$ (blue), $\hbar=2\pi$ (orange), and $\hbar=\pi$ (green).
Note that the horizontal axis is $N^{3/2}$.
The dots are the exact values of free energy at integer $N$,
while
the solid curves represent 
the perturbative free energy
$F_\text{pert}=-\log Z_\text{pert}(N,\hbar)$ given by the Airy function \eqref{zpert}.
}
  \label{fig:Zpert}
\end{figure}

One can compare all these found values with what is predicted by the conjecture of \cite{GHM}. In order to proceed, we must construct another spectral invariant out of the traces, which we call the ``fermonic spectral traces'' $Z(N)$. These are just the $N^{\rm th}$ coefficients of the small $\kappa$ expansion of the Fredholm determinant $\det (1+\kappa \rho )$, and can be computed from the traces as
\be
	\label{fermfromtrace}
	Z(N)=\sum_{\text{partitions} \, \{m_\ell\}} \!\!\!\!\!\!\!\!\!\!' \quad \prod_\ell \frac{(-1)^{(\ell-1)m_\ell}}{m_\ell ! \, \ell^{m_\ell}} (\tr \rho^\ell)^{m_\ell},
\ee
where the sum is over the integer partitions $\{ m_\ell\}$ of $N$:
\be
	\sum_\ell \ell m_\ell=N.
\ee
One of the consequences of the proposal in \cite{GHM} is that for a given $\hbar$, this number is given by
\be
	\label{airyZ}
	Z(N,\hbar)=\frac{1}{2\pi \ri} \int_{\gamma} \re^{J(\mu,\hbar)-N \mu}.
\ee
In this expression, $\gamma$ is the contour that is used to define the Airy function, and $J(\mu,\hbar)$ is the so-called modified grand-potential. This last quantity is conjectured in \cite{HMMO,GHM} to be constructed in a very precise way from the free-energy of the refined topological string in the Nekrasov-Shatashvili limit and the free-energy of the standard topological string. 
When $\hbar$ is $\pi$ times a rational number,
these two contributions have a pole at that value of $\hbar$, 
but as shown in \cite{HMMO} the sum of them are actually finite in total, thanks to the
pole cancellation between worldsheet instantons and membrane instantons \cite{HMOinst}. 
The outcome of all this, is that for a fixed (finite !) value of $\hbar$, we can express $J(\mu,\hbar)$ as
\be
	\label{Jmod}
	J(\mu,\hbar)=\frac{C(\hbar)}{3}\mu^3+B(\hbar)\mu+A(\hbar)+\sum_{d>0} p_d(\mu)\re^{-d\mu}.
\ee
In the above expression, the polynomial part in $\mu$ actually corresponds to $J^{(\rm p)}(\mu)$ given in (\ref{Jmuparts}). The sum is over a countable set ${d}$ of ``instanton levels'' which are not necessarily integers, and $p_d(\mu,\hbar)$ is a quadratic polynomial in $\mu$ with  known coefficients, which are related through the conjecture of \cite{GHM} to some enumerative invariants of local $\mathbb P^2$ (the $N^d_{j_L,j_R}$ of section 2.3). 
Plugging \eqref{Jmod} into \eqref{airyZ} and expanding the instanton part, we find
that $Z(N,\hbar)$ is written as
\begin{equation}
 Z(N,\hbar)=Z_\text{pert}(N,\hbar)+Z_\text{np}(N,\hbar),
\end{equation}
where $Z_\text{pert}(N,\hbar)$ is the perturbative
part given by the Airy function \cite{MP}
\begin{equation}
Z_\text{pert}(N,\hbar)=C(\hbar)^{-\frac{1}{3}}e^{A(\hbar)}\text{Ai}\Big[C(\hbar)^{-\frac{1}{3}}(N-B(\hbar))\Big], 
\label{zpert}
\end{equation}
while $Z_\text{np}(N,\hbar)$ is the non-perturbative part  
given by some combination of the derivative of Airy functions \cite{HMOinst}.
As we can see in Figure \ref{fig:Zpert},
the exact values of free energy
$F=-\log Z(N,\hbar)$ exhibit a nice agreement with the
perturbative free energy
$F_\text{pert}=-\log Z_\text{pert}(N,\hbar)$ given by the Airy function \eqref{zpert}.
Since the non-perturbative part $Z_\text{np}(N,\hbar)$ rapidly decreases at large $N$, 
we have a converging approximation to the exact value of the partition function
by the expansion in terms of the instanton level $d$. 
Using this technique, we can compare these converging approximations to the exact values obtained above. We find a perfect agreement in all cases, even at small $N$.

As an example, let us see how it works for the traces in the case of $\hbar=6\pi$. The TBA system produces the non-trivial looking numbers given in (\ref{6pitraces}), from which it is straightforward to construct the exact fermonic spectral traces using (\ref{fermfromtrace}). From the conjecture side, we first compute the modified grand potential $J(\mu,\hbar)$. For 
$\hbar=6\pi$, it can actually be done exactly (see results in appendix \ref{app3}), and its large $\mu$ expansion is given by
\be
\ba
	\label{Jmu6pi}
	J(\mu,6\pi) &=\left(\frac{\mu ^3}{8 \pi ^2}-\frac{7 \mu }{24}+A(6\pi) \right)-\re^{-\mu}-\frac{3 \re^{-2\mu}}{2}+\frac{\left(-15 \mu ^2-6 \mu +33 \pi ^2-2\right)
    \re^{-3\mu}}{8 \pi ^2}\\
    & \qquad  -\frac{27  \re^{-4\mu}}{4}-\frac{81  \re^{-5\mu}}{5}+\frac{3 \left(-222 \mu ^2-14 \mu +642 \pi ^2-5\right)  \re^{-6\mu}}{32 \pi
   ^2} \\
   & \qquad -\frac{729 \re^{-7 \mu}}{7}-\frac{2187 \re^{-8 \mu}}{8}+ O\left( \re^{-9\mu}\right).
\ea
\ee
The value of $A(6\pi)$ can be written down in exact form using results of \cite{GHM,HO14}:
\be
A(6\pi)=\frac{1}{9} \log \left(2^{21} \sin ^4\left(\frac{\pi }{9}\right) \sin ^7\left(\frac{2
   \pi }{9}\right) \cos ^{10}\left(\frac{\pi }{18}\right)\right)-\frac{\zeta (3)}{9\pi
   ^2}.
\ee
With the expression of $J(\mu,6\pi)$ in (\ref{Jmu6pi}) up to instanton correction $\re^{-d\mu}$, we compute the numerical predictions for the fermonic spectral traces (\ref{airyZ}).
Table \ref{table1} shows the very good agreement between the numbers, and also the fast convergence when one increases the value of $d$.
%
%
\begin{table}[htb]
\begin{center}
\begin{tabular}{ | c |  l |}
	\hline
  up to $\re^{-d\mu}$ with & $Z(0,6\pi) $ \\ \hline
  $d=0$ 			  & {\bf 1.00} 841... \\
  $d=2$ 			  & {\bf 1.0000000} 160... \\ 
  $d=4$ 			  & {\bf 0.99999999999999999} 028... \\
  $d=6$ 		   	  & {\bf 1.000000000000000000000000000} 394... \\
  $d=8$			  & {\bf 1.000000000000000000000000000000000000000} 352... \\ \hline
  exact value 	          & {\bf 1} \\
  \hline
	\hline
  up to $\re^{-d\mu}$ with & $Z(1,6\pi) \cdot 10^{3} $ \\ \hline
  $d=0$ 			  & {\bf 8.} 403... \\
  $d=2$ 			  & {\bf 8.39561470} 120... \\ 
  $d=4$ 			  & {\bf 8.395614700210276511} 486... \\
  $d=6$ 		   	  & {\bf 8.39561470021027651154265956252} 243... \\
  $d=8$			  & {\bf 8.395614700210276511542659562521945624932134} 290... \\ \hline
  exact value 	          & {\bf 8.39561470021027651154265956252194562493213418137} \\
  \hline
  	\hline
  up to $\re^{-d\mu}$ with & $Z(2,6\pi) \cdot 10^{6} $ \\ \hline
  $d=0$ 			  & {\bf 7.81} 301... \\
  $d=2$ 			  & {\bf 7.8116449158} 766... \\ 
  $d=4$ 			  & {\bf 7.811644915860187507372} 177... \\
  $d=6$ 		   	  & {\bf 7.811644915860187507372286933587324} 472... \\
  $d=8$			  & {\bf 7.8116449158601875073722869335873242743359331556} 428... \\ \hline
  exact value 	          & {\bf 7.811644915860187507372286933587324274335933155632305} \\
  \hline
\end{tabular}
\normalsize
\end{center} 
\caption{ \label{table1} Here are the first two fermonic spectral traces and the normalisation check for $\hbar=6\pi$ against the numerics generated using the conjecture. Each inclusion of the next two instanton sectors improves the numerical results by more than 10 digits.}
\end{table}

By including $25$ levels of instanton correction, we can reproduce around 200 digits of the exact expressions.

  						%
 \section{The TBA system for more general $(m,n)-$operators}
\label{sec:TBA-mn} 
 
In this section, we propose a generalization for the construction of the TBA system of local $\mathbb P^2$ to the more general family of kernels of the inverse $(m,n)-$operator 
 \be
 	{\rm O}_{m,n} = \re^{\rm x}+\re^{\rm y}+\re^{-m {\rm x} -n\rm y},
 	\qquad \qquad {\rho}_{m,n} = {\rm O}^{-1}_{m,n}.
 \ee
 where we take $m,n \in \mathbb Z_{\geq 0}$.
These are the quantized mirror curves of various blow-downs of local del Pezzo Calabi-Yau threefolds, so they are interesting to study as they form the building blocks of more complicated geometries. They can be viewed as the mirrors of the total space of the anticanonical bundle over the weighted projective space $\mathbb P(1,m,n)$.
All the $ {\rho}_{m,n}$ are of trace class on $L^2(\mathbb R)$, and can be factorized as \cite{KM} 
  \be
   	\label{mnOp}
 	{\rho}_{m,n} =W({\rm p}) \frac{\re^{ C\rm q} }{2 \cosh \left ( \frac{1}{2}\rm q \right )} W({\rm p})^*,
 \ee
The exact linear relation between $\rm x, y$ and $\rm q,p$ can be found in \cite{KM}, suffice to say their commutator is $[{\rm q,p}]=\ri h \equiv \ri(m+n+1)\hbar$. Also, the number $C$ and the function $W(p)$ are given by
 \be
 \ba
 	C &=\frac{m-n+1}{2(m+n+1)}, \\
 	W({\rm p}) &=\re^{\frac{m}{2(m+n+1)}  {\rm p}} \Phi_b\left (\frac{{\rm p} +\ri \hbar \frac{m+1}{2}}{2 \pi b} \right ), \qquad \qquad b^2= \frac{(m+n+1) \hbar}{2\pi}.
\ea
 \ee
 We will see that the value of $C$ will be critical in the construction of the TBA, whereas the exact form of $W(p)$ will not influence its structure. As in section \ref{sec:TBA-P2}, we rather study the similarity-related operator 
  \be
  	\label{mnOpMOD}
 	{\rm \rho}=\sqrt{V(\rm p)}  \frac{\re^{C \rm q} }{2 \cosh \left ( \frac{1}{2}\rm q \right )}  \sqrt{V(\rm p)}, \qquad \qquad V({\rm p})=W({\rm p})W({\rm p})^*,
\ee
which has the same spectrum as the initial operator.
The integral kernel of $\rho$ in the $p$ representation can be readily found from (\ref{mnOpMOD}):
   \be
   	\label{mnkernel}
 	{\rho}(p,p')= \frac{\sqrt{V(p)V(p')}}{2 h \cosh \left ( \frac{\pi}{h}(p-p')+\ri \pi C \right )} = \frac{\sqrt{E(p) E(p')}}{\alpha M(p) +\alpha^{-1} M(p')},
 \ee
with
  \be
  \ba
 	\alpha & =\re^{ \ri \pi C }, \qquad \qquad \qquad \qquad \omega \equiv -\alpha^{-2}=\re^{2 \pi \ri \frac{n}{m+n+1}},\\
	E(p)&=\frac{1}{h}M(p) V(p), \qquad \quad M(p) =\re^{\frac{2\pi}{h}p}.
\ea
 \ee
As promised in the introduction, this kernel has the shifted $\cosh$ structure (\ref{generalisedkernel}). Again, we will consider a more general case by letting the $V(p)$ function be an arbitrary real function with good analytic properties, since the TBA system will only depend trivially on it. With these definitions, it is easily seen that (\ref{phijdef}) and (\ref{powerkernel}) are valid as they stand. Moreover, we define the resolvent operator as (\ref{resolvent}), which is related to the grand-potential through (\ref{RtoJ}).
 
 We saw in the particular case of the local $\mathbb P^2$ kernel that it is the nature of $\omega$ which determines the structure of the TBA equations. In the local $\mathbb P^2$ case, it was a non-trivial $3^{\rm rd}$ root of 1, which implied a splitting of the resolvent in three parts, every part having a definite ``charge'' under this $\mathbb Z_3$ group. In the general case, we see that $\omega$ is a $D^{\rm th}$ root of unity, where $D$ is the very important quantity given by 
 \be
 D=\frac{m+n+1}{ {\rm gcd}(n,n+m+1)} \in \mathbb N_{> 0},
 \ee
 and the splitting is expected to be according to the $\mathbb Z_D$ cyclic group. It must be stressed that our construction works only if $C$ is a rational number. The case where $C$ is a non-rational real number or even a complex number is certainly of a more complicated nature which we shall not consider.
 The resolvents $\rm R_k$ (which have definite $k$-``charge'' under $\mathbb Z_D$) now become
 \be
 	\label{RkfromRD}
	{\rm R}_k(\kappa)=\frac{1}{D}\sum_{r=0}^{D-1} \omega^{-r k} {\rm R}(\omega^r \kappa)=\frac{\kappa^k \rho^{k+1}}{1-(\kappa \rho)^D}, \qquad \qquad (k=0,1,...,D-1)
\ee
 	and all the definitions and relations (\ref{Rkrelations})-(\ref{limofkResolvent}) go through, with the obvious modification that the labels $k$ and $r$ in $R_k$ and $\Phi_r$ run from $0$ to $D-1$. We then find the following generalizations of relations (\ref{zerocomb}) and (\ref{nonvanishingRcomb}) :
 \be
 	\label{zerocombD}
 	\sum_{r=0}^{D-1} \omega^r \Phi_{r}(p) \bar \Phi_{D-1-r}(p) =0,
 \ee
and 
 \be
\ba
	\label{nonvanishingRcombD}
	R_k(p)&=\frac{V(p)}{h} \frac{1}{2\sin \left ( \pi\frac{(k+1)n}{m+n+1} \right )}\sum_{r=0}^{D-1} \omega^{r-\frac{k}{2}} \Phi_{r}(p) \bar \Phi_{k-r}(p), \qquad \qquad (k = 0,1,...,D-2), \\
	R_{D-1}(p)&=-\frac{V(p)}{4\pi} \sum_{r=0}^{D-1}\omega^{r+\frac{1}{2}} \{\Phi_r(p),\bar \Phi_{D-1-r}(p) \},
\ea
\ee
where the labels of $\Phi_k$ and its conjugate should be understood mod $D$. The linear relations and the quantum Wronskian relations are generalized as
\be
\ba
	\label{PhidifferenceD}
	\Phi_r^{+n}-\Phi_r^{-(m+1)} &=\kappa \omega^{-1/2}V \Phi_{r-1}, \\
	\bar \Phi_r^{-n}-\bar \Phi_r^{+(m+1)} &=\kappa \omega^{1/2}V \bar \Phi_{r-1},
\ea
\ee
and
\be
\ba
	\sum_{r=0}^{D-1} \omega^r \Phi_r^{+\frac{n}{2}} \bar \Phi_{D-r}^{-\frac{n}{2}}=1, \qquad \qquad \qquad
	\sum_{r=0}^{D-1} \omega^r \Phi_r^{-\frac{m+1}{2}} \bar \Phi_{D-r}^{+\frac{m+1}{2}}=1.
\ea
\ee
Interestingly, we need in general two quantum Wronskian relations, whereas in the prototypical $C=0$ case proven in \cite{TW}, there is only one such relation. This is compatible since the $C=0$ case corresponds to $(m,n)=(n-1,n)$, in which case $\Phi_r$ are real, and thus the two quantum Wronskians given above coincide. The auxiliary functions $\eta_k$ and $\tilde \eta_k$ for $k=0,1,...,D-2$ are defined as
\be
\ba
	\eta_k&=\omega^{\frac{D}{2}} \sum_{r=0}^{D-1} \omega^{r-\frac{k+1}{2}} \Phi_r^{+\frac{n}{2} } \bar \Phi_{k+1-r}^{-\frac{n}{2} }, \\
	\tilde \eta_k&=\omega^{\frac{D}{2}} \sum_{r=0}^{D-1} \omega^{r-\frac{k+1}{2}} \Phi_r^{-\frac{m+1}{2} } \bar \Phi_{k+1-r}^{+\frac{m+1}{2} },
\ea
\ee
and satisfy the following relations
\be 
\ba
	\label{mainetaD}
	\eta_k^{+\frac{n}{2}}-\tilde \eta_k^{-\frac{m+1}{2}} &= \eta_k^{-\frac{n}{2}}-\tilde \eta_k^{+\frac{m+1}{2}}=2\omega^{\frac{D}{2}} h \kappa \sin \left ( \frac{(k+1)n}{m+n+1}\pi\right ) R_k, \\
	\eta_k &= -\frac{\sin \left (\frac{m+1}{2} \hbar \partial_p \right )}{ \sin \left (\frac{n}{2} \hbar \partial_p \right ) }\tilde \eta_k.
	\ea
\ee
Now let us give the appropriate generalization of $\Psi$, $\mathcal R$ and $\mathcal H$. This will depend on the parity of the integer $D$ (whether it is even or odd). The issue arises in the rearranging of the double sum when one adapts the steps of (\ref{Rmanip}). We will see that the following definitions work:
\be
\ba
	\Psi=\sum_{r=0}^{D-1}\omega^{- \frac{D-1}{2} r+\frac{\delta}{2} r} \Phi_r, \qquad \qquad \tilde  \Psi=\sum_{r=0}^{D-1} \omega^{ \frac{D-1}{2} r+\frac{\delta}{2} r} \tilde  \Phi_r.
\ea
\ee
where
\be
\ba
	\delta=
	\begin{cases}
		0 & \qquad D \text{ is odd} \\
		1 & \qquad D \text{ is even}. \\
	\end{cases}
\ea
\ee
The $\mathcal R$ becomes
\be
\ba
	\mathcal R &= \Psi \tilde \Psi =\sum_{r,s=0}^{D-1} \omega^{ \frac{D-1}{2}(s-r)+\frac{\delta}{2}(s+r) } \Phi_r \bar \Phi_s.
\ea
\ee
 We now rearrange the terms in the sum using $\ell=r$, $k=s+r$. But we must be careful: in order to take the sum over $\ell,k=0,1,...,D-1$, we need that each term (or at least those with  $s+r \geq D$) be invariant under $s \rightarrow s+D$. Under such a shift, the $\bar \Phi_s$ functions stay the same because their label is defined modulo $D$, and the power of $\omega$ acquires an extra $ \omega^{D(D-1+\delta)/2}$. But since we can always find an integer $n$ such that $D=2n+1-\delta$ (thanks to our definition of $\delta$), the extra factor reduces to $\omega^{nD}=1$. So we are allowed to write
 \be
\ba
	\mathcal R &= \sum_{\ell,k=0}^{D-1} \omega^{ \frac{D-1}{2}(k-2\ell)+\frac{\delta}{2}k } \Phi_\ell \bar \Phi_{k-\ell}.  \\
			& = \sum_{k=0}^{D-2} \omega^{\frac{D+\delta}{2}k} \frac{2 h}{V} \sin \left (\frac{(k+1)n}{m+n+1} \pi  \right ) R_k,
\ea
\ee
where in the last line we used (\ref{zerocombD})-(\ref{nonvanishingRcombD}). We take as our $\mathcal H$ auxiliary function the following:
\be
	\label{Hdef}
	\mathcal H = \frac{\sin \left ( \frac{1}{2}\hbar \partial_p \right )}{ \sin \left ( \frac{n}{2}\hbar \partial_p \right ) }  \sum_{k=0}^{D-2} \omega^{\frac{D+\delta}{2}k}  \tilde \eta_k,
\ee
and with the use of (\ref{mainetaD}), we arrive at our first TBA relation (which has the same form for odd and even $D$) :
\be
	\label{mntba1}
\ba
	-\omega^{\frac{D}{2}} \kappa V \mathcal R &= \sum_{s=-\frac{m+n}{2}}^{\frac{m+n}{2}} \mathcal H^{+s}.
\ea
\ee
The counting index $s$ is incremented by one in the above sum. Notice that $\omega^{D/2}$ is $\pm 1$ depending on the value of $m,n$. For the second TBA equation, let us define for convenience
\be
\ba
	 \mathcal G_1 &\equiv  \Psi^{+\frac{n}{2}} \tilde \Psi^{-\frac{n}{2}}=1+\sum_{k=0}^{D-2} \omega^{\frac{D}{2}k+\frac{\delta}{2}(k+1)} \eta_k=1-\omega^{\frac{\delta}{2}} \sum_{s=-\frac{m}{2}}^{\frac{m}{2}} \mathcal H^{+s}, \\
	 \mathcal G_2 &\equiv \Psi^{-\frac{m+1}{2}} \tilde \Psi^{+\frac{m+1}{2}}=1+\sum_{k=0}^{D-2} \omega^{\frac{D}{2}k+\frac{\delta}{2}(k+1)} \tilde \eta_k=1+\omega^{\frac{\delta}{2}} \sum_{s=-\frac{n-1}{2}}^{\frac{n-1}{2}} \mathcal H^{+s},
\ea
\ee
(here we used the quantum Wronskian relations). As in the local $\mathbb P^2$ case, we get the second TBA equation by shifting appropriately the above terms :
\be
	\label{mntba2}
	\prod_{s=0}^{\frac{\rm lcm}{n}} \mathcal R^{-\frac{\rm lcm}{2}+ns} \prod_{s=1}^{\frac{\rm lcm}{m+1}-1} \mathcal R^{-\frac{\rm lcm}{2}+(m+1)s}= \prod_{s=0}^{\frac{\rm lcm}{n}-1} \mathcal G_1^{-\frac{\rm lcm}{2}+n(s+\frac{1}{2}) } \prod_{s=0}^{\frac{\rm lcm}{m+1}-1} \mathcal G_2^{-\frac{\rm lcm}{2}+(m+1)(s+\frac{1}{2}) },
\ee
where
\be
	 {\rm lcm}={\rm lcm}(n,m+1)
\ee
is the lowest common multiple of the integers $n$ and $m+1$.
Finally, in the same way as for the special case of local $\mathbb P^2$, we define
\be
\ba
	\mathcal W & =\{ \Psi, \tilde \Psi\}=\ri \left ( \Psi' \tilde \Psi - \Psi  \tilde \Psi' \right ).
\ea
\ee
By a calculation similar to (\ref{3rdtba}) which involves appropriate shifts of the $\mathcal R$ and the $\mathcal G_i$, we find that the last TBA equation is given by
\be
	\label{mntba3}
\ba
	\left ( \frac{\mathcal W}{\mathcal R} \right )^{+\frac{\rm{lcm}}{2}} -\left (  \frac{\mathcal W}{\mathcal R} \right )^{-\frac{\rm{lcm}}{2}}
	=\ri \frac{\partial}{\partial p} \log \left (
	\frac{
	\prod_{s=0}^{\frac{\rm lcm}{n}-1} \mathcal G_1^{-\frac{\rm lcm}{2}+n(s+\frac{1}{2}) } \quad
	\prod_{s=1}^{\frac{\rm lcm}{m+1}-1} \mathcal R^{-\frac{\rm lcm}{2}+(m+1)s}
	}{
	 \prod_{s=0}^{\frac{\rm lcm}{m+1}-1} \mathcal G_2^{-\frac{\rm lcm}{2}+(m+1)(s+\frac{1}{2}) } \quad
	 \prod_{s=1}^{\frac{\rm lcm}{n}-1} \mathcal R^{-\frac{\rm lcm}{2}+ns}
	}
	 \right ).
\ea
\ee
Equations (\ref{mntba1}), (\ref{mntba2}) and (\ref{mntba3}) form the difference TBA system for the $(m,n)$-operator. From the solution $(\mathcal R, \mathcal H, \mathcal W)$, we can extract the diagonal resolvents as
\be
\ba
	R_k(\kappa)&=\frac{V\omega^{-\frac{D+\delta}{2}k}}{2 (m+n+1) \hbar D \sin \left ( \frac{(k+1)n}{m+n+1}\pi \right )} \sum_{r=0}^{D-1} \omega^{-r k} \mathcal R(\omega^r \kappa), \qquad \qquad (k=0,1,...,D-2) \\
	R_{D-1}(\kappa)&=-\frac{V \omega^{-\frac{D(D-2)+\delta(D-1)}{2}}}{4 \pi D} \sum_{r=0}^{D-1} \omega^{r} \mathcal W(\omega^r \kappa).
\ea
\ee

The equations forming the TBA system can be inverted to their integral version, providing the generalization of (\ref{p2TBAint}). However, it happens that the difference TBA system (\ref{mntba1}), (\ref{mntba2}) and (\ref{mntba3}) is not always written in its optimal form, and can sometimes be simplified by a redefinition of the auxiliary function $\mathcal H$, leading to simpler integral systems when inverted. For example, the system (\ref{mntba1}), (\ref{mntba2}) and (\ref{mntba3})  for the case $C=0$, corresponding to $(m,n)=(1,2)$, yields
\be
\ba
	\kappa V \mathcal R&=\mathcal H^{-\frac{3}{2}}+\mathcal H^{-\frac{1}{2}}+\mathcal H^{\frac{1}{2}}+\mathcal H^{-\frac{3}{2}}, \\
	\mathcal R^{-1} \mathcal R^{+1}&=\left (1-\ri \mathcal H^{-\frac{1}{2}}-\ri \mathcal H^{+\frac{1}{2}} \right )\left (1+\ri \mathcal H^{-\frac{1}{2}}+\ri \mathcal H^{+\frac{1}{2}} \right ), \\
	\left ( \frac{\mathcal W}{\mathcal R} \right )^{+1} -\left (  \frac{\mathcal W}{\mathcal R} \right )^{-1}
	&=\ri \frac{\partial}{\partial p} \log \left ( \frac{1-\ri \mathcal H^{-\frac{1}{2}}-\ri \mathcal H^{+\frac{1}{2}}}{1+\ri \mathcal H^{-\frac{1}{2}}+\ri \mathcal H^{+\frac{1}{2}}} \right ).
\ea
\ee
This can be simplified by introducing $\eta=\mathcal H^{-\frac{1}{2}}+\mathcal H^{+\frac{1}{2}}$, so that we recover the following difference form of the $C=0$ case: 
\be
\ba
	\kappa V \mathcal R&=\eta^{-1}+\eta^{+1},\\
	\mathcal R^{-1} \mathcal R^{+1}&=1+\eta^2, \\
	\left ( \frac{\mathcal W}{\mathcal R} \right )^{+1} -\left (  \frac{\mathcal W}{\mathcal R} \right )^{-1}
	&=\frac{2 \eta'}{1+\eta^2},
\ea
\ee
which lead to the standard TBA integral system for this case (see for example eq. (4.10) and (4.11) of \cite{CM}).
So in general (\ref{mntba1}), (\ref{mntba2}) and (\ref{mntba3}) are expected to be true, but may sometimes be further simplified.

We can repeat the similar analysis for $\rho_{m,n}$
as in the local $\mathbb P^2$
case, namely the semi-classical expansion and the exact computation of the spectral traces.
For instance, the exact values of 
the ``fermionic spectral trace'' $Z_{m,n}(N,\hbar)$ for the $(m,n,\hbar)=(3,1,2\pi)$
case are given by
\begin{equation}
\begin{aligned}
 Z_{3,1}(1,2\pi)&=\frac{1}{5\sqrt{5}},\quad
Z_{3,1}(2,2\pi)=\frac{\sqrt{5}-2}{125}, \\
Z_{3,1}(3,2\pi)&=-\frac{\sqrt{\frac{1}{10} \left(5+\sqrt{5}\right)}}{60 \pi
   }+\frac{35-3 \sqrt{5}}{6250},\\
Z_{3,1}(4,2\pi)&=\frac{\sqrt{\frac{1}{2} \left(205+89 \sqrt{5}\right)}}{3000 \pi }-\frac{9
   \left(3+\sqrt{5}\right)}{31250},\\
Z_{3,1}(5,2\pi)&=-\frac{1}{800 \pi
   ^2}-\frac{\sqrt{377+\frac{622}{\sqrt{5}}}}{30000 \pi }+\frac{2175-277 \sqrt{5}}{3906250},\\
Z_{3,1}(6,2\pi)&=-\frac{19}{36000 \sqrt{5}
   \pi ^2}-\frac{\sqrt{13505+5038 \sqrt{5}}}{750000 \pi }+\frac{495 \sqrt{5}-221}{9765625},\\
Z_{3,1}(7,2\pi)&=
-\frac{53
   \sqrt{5}-91}{1800000 \pi
   ^2}-\frac{\sqrt{9841-\frac{21982}{\sqrt{5}}}}{3750000 \pi
   }+\frac{7949 \sqrt{5}-16885}{488281250}.
\end{aligned} 
\end{equation}
They are consistent with the conjecture of the fermionic spectral trace of resolved
$\mathbb C^3/\mathbb Z_5$ orbifold \cite{Codesido:2015dia}.\footnote{We would like to thank Marcos Mari\~no for discussion on this point.}

  						%
 \section{Conclusion}
\label{sec:conclusion} 
 
 In this paper, we have generalized the correspondence between the spectral functions of some kernels and the solutions of some non-linear TBA-like systems of equations. The prototypical example proposed in \cite{CFIV,Zam} and proved in \cite{TW} acted as a beacon for our investigation, and our result is a generalization of that case. Let us emphasize the two most important features of the generalization, which are the decomposition of the resolvent function according to the $\mathbb{Z}_D$
symmetry, and the generalized quantum Wronskian relations. Our study mainly focused on another specialization of this construction: the case which is built from the quantized mirror curve of local $\mathbb P^2$. We used our newly found TBA-like system to compute some traces of the local $\mathbb P^2$ kernel. We also expanded the TBA system semiclassically, which proved to be a good substitute for WKB methods in the semiclassical investigation of the operator. This new TBA system proved to be a useful tool in order to do some further testings of the proposal given in \cite{GHM} (and nicely reviewed in \cite{Mreview}).

It is clear that the precise structure of the kernel was very important for extracting a TBA-like system of it. In all our cases it involved the $1/\cosh$ function. It would be interesting to see if there can be a similar relation between TBA-like integral equations and kernels having other forms. Also the more general story where we allow non rational shifts in the $\cosh$ function might be interesting, even if it is not crucial in the view of the examples coming from topological strings and mirror symmetry. It is also important to mention that in the $(m,n)$ case, the mirror curve underlying the operator has higher genus. In fact, we could introduce many deformation parameters which would result in more complicated operators. Our TBA system only includes a single parameter (whose role is played by $\kappa$), while the others are set to 0.  Finding generalizations including more parameters for both the operator kernel and its related TBA would be a very interesting avenue of further study. Finally, investigating new techniques for solving more explicitly the TBA-like equations would also be of interest, especially if it helps to understand the behavior of the resolvent 
when both $\kappa$ and $\hbar$ become large (with the appropriate scaling). 
Indeed, this limit is of particular interest in the light of the proposal of \cite{GHM}, since it is the one that probes the non-perturbative completion of the grand potential. It is the ``strong quantum coupling regime'' of the operator $\rho$, which is typically harder to probe.

\vskip6mm 
\acknowledgments
 The authors are grateful to Yasuyuki Hatsuda and Marcos Mari\~no for useful discussions.
KO would like to thank the theory group in the University of Geneva for hospitality.
This research was partly supported by the NCCR SwissMAP, funded by the Swiss National Science Foundation.
 
  						%
 \appendix
 
\section{Kernel of $\rho^\ell$} \label{apptracerho}
In this Appendix, we explain how to obtain the expression (\ref{powerkernel}) for the kernel of $\rho^\ell$.\footnote{A similar computation has appeared in a recent paper \cite{Nosaka:2015iiw}.}
The kernel of $\rho$ is given by (\ref{rhofullkernel}). For convenience, we switch from integral operator notation to a more abstract operator notation. Introducing the operators $M$ and $\sqrt{E} \otimes \sqrt{E}$ with kernels
\be
\ba
	M(p,p')=M(p)\delta(p-p'), \qquad \qquad (\sqrt{E} \otimes \sqrt{E})(p,p')=\sqrt{E(p)E(p')},
\ea
\ee
we can write $ (\ref{rhofullkernel})$ as
\be
\ba
	\alpha M \rho +\alpha^{-1} \rho M=\sqrt{E}\otimes \sqrt{E}.
\ea
\ee
Let us introduce the $\alpha$-twisted commutators and anti-commutators
\begin{align}
	[A,B]_{\alpha}=\alpha AB-\alpha^{-1} BA, \qquad \qquad \{A,B\}_{\alpha}=\alpha AB+\alpha^{-1} BA,
\end{align}
which satisfy
\begin{align} \nonumber
	[A,BC]_{\alpha^n}=\alpha \{A,B\}_{\alpha^{n-1}}C-\alpha^{-n+1} B \{A,C\}_{\alpha}, \\	
	\{A,BC\}_{\alpha^n}=\alpha [A,B]_{\alpha^{n-1}}C+\alpha^{-n+1} B \{A,C\}_{\alpha}.
\end{align}
Then it is easily seen that
\begin{align} \nonumber
	\{M,\rho\}_\alpha &= \sqrt{E} \otimes \sqrt{E}, \\  \nonumber
	[M,\rho^2 ]_{\alpha^2} &= \sqrt{E} \otimes \sqrt{E}\alpha \rho-\alpha^{-1} \rho \sqrt{E} \otimes \sqrt{E}, \\  \nonumber
	\{ M,\rho^3 \}_{\alpha^3} &= \sqrt{E} \otimes \sqrt{E}\alpha^2 \rho^2-\alpha^{-1} \rho \sqrt{E} \otimes \sqrt{E} \alpha \rho+\alpha^{-2} \rho^2 \sqrt{E} \otimes \sqrt{E}, \\
	...
\end{align}
and in general
\begin{align} 
	\label{rhoelleq1}
	\alpha^\ell M \rho^\ell-(-1)^\ell \alpha^{-\ell} \rho^\ell M=\alpha^{\ell-1}\sum(-\alpha^{-2})^k \rho^k \sqrt{E} \otimes \sqrt{E} \rho^{\ell-k-1}. 
\end{align}
We now write everything in terms of kernels. To do that, it is convenient to define
\be
	\phi_k(p) \equiv \frac{1}{\sqrt{E(p)}} (\rho^k \sqrt{E})(p), \qquad \text{so} \qquad \bar \phi_k(p) = \frac{1}{\sqrt{E(p)}} (\sqrt{E} \rho^k)(p),
\ee
which is nothing else than (\ref{phijdef}) (for the second equality above, we use the hermiticity of the operator $\rho$). Using these functions and $-\alpha^{-2}=\omega$, (\ref{rhoelleq1}) becomes
\be
	M(p)\rho^\ell(p,p')-\omega^\ell \rho^\ell(p,p')M(p')=\alpha^{-1} \sqrt{E(p)E(p')}\sum \omega^k  \phi_{k}(p) \bar \phi_{\ell-k-1}(p').
\ee
The result (\ref{powerkernel}) readily follows.

\section{Quantum Wronskians} \label{appQW}
We give here a motivation for the quantum Wronskian relations for the local $\mathbb P^2$ case. The generalization to the $(m,n)$ case is straightforward.
Let us define
\be
\ba
	I&=\sum_{r=0}^{2} \omega^r \Phi_r^{+\frac{1}{2}} \bar \Phi_{3-r}^{-\frac{1}{2}}-1, \qquad \qquad
	\tilde I& = \sum_{r=0}^{2} \omega^r \Phi_r^{-1} \bar \Phi_{3-r}^{+1}-1.
\ea
\ee
The quantum Wronskians are true if $I=\tilde I=0$. Using relations (\ref{Phidifference}) and (\ref{zerocomb}), we deduce that
\be
\ba
	\label{Irel1}
	I^{+\frac{1}{2}}- \tilde I^{-1}=I^{-\frac{1}{2}}-\tilde I^{+1}=0,
\ea
\ee
which implies
\be
\ba
	\sin \left ( \frac{  \hbar}{2} \partial_p \right ) I=-\sin \left (  \hbar \partial_p \right )\tilde I.
\ea
\ee
If we formally denote the inverse of the shift operator $\sin(\alpha \partial_p)$ by $\frac{1}{\sin(\alpha \partial_p)}$, we get
\be
\ba
	\label{Irel2}
	 I=- \frac{ \sin \left (  \hbar \partial_p \right ) }{\sin \left ( \frac{  \hbar}{2} \partial_p \right )}\tilde I.
\ea
\ee
We assume that such manipulations are well defined, as the functions $\Phi_r \bar \Phi_{3-r} \,\, (r >0)$ and $\Phi_0 \bar \Phi_{0}-1$ have the right analytic properties (at all orders in $\kappa$). 
Plugging (\ref{Irel2}) into (\ref{Irel1}), we get 
\be
\ba
	\frac{\sin \left (\frac{3}{2}  \hbar \partial_p \right )}{\sin \left ( \frac{1}{2}  \hbar \partial_p \right )} \tilde I=0, \qquad  \frac{\sin \left (\frac{3}{2}  \hbar \partial_p \right )}{\sin \left ( \hbar \partial_p \right )}  I=0,
\ea
\ee
from which we conclude that we should have $\tilde I = I =0$. The rigorous version goes certainly along the lines of \cite{TW}.

\section{Some results for local $\mathbb P^2$ when $\hbar=6\pi$}
\label{app3}

In the case $\hbar = 6\pi$, the modified grand potential for local $\mathbb P^2$ can be found exactly. It is given by
 \begin{equation}
 	J(\mu,6\pi)=A(6\pi)-\frac{7}{24}\mu+\frac{1}{12 \pi^2}\left ( F_0- t \partial_t F_0+\frac{t^2}{2} \partial_t^2 F_0 \right )+f_1,
 \end{equation}
 where
  \begin{equation}
  \begin{aligned}
 	t &=3\mu+w_1, \\
	\partial_t F_0 &= \frac{1}{6}\left ( t^2-w_1^2+w_2 \right ) \equiv \frac{t^2}{6}+\partial_t F_0^{\rm inst},
\end{aligned}
 \end{equation}
 and 
  \begin{equation}
  \begin{aligned}
 	w_1 &=-3 \sum_{\ell=1}^{\infty} \frac{(3\ell-1)!}{(\ell !)^3} \re^{-3\ell \mu}=-6\re^{-3\mu} \,_4 F_3 \left ( 1,1,\frac{4}{3},\frac{5}{3}; 2,2,2; 27\re^{-3\mu} \right ) \, \\
	w_2 &=18 \sum_{\ell=1}^{\infty} \frac{(3\ell-1)!}{(\ell !)^3} \left [ \psi(3\ell)-\psi(\ell+1) \right ] \re^{-3\ell \mu}, \\
	f_1 &= \frac{1}{3}\log(1-3\re^{-\mu})-\frac{3}{8}\log (1-27\re^{-3\mu})-\frac{1}{2}\log \left (\frac{\rd t}{\rd (3\mu)} \right ) .
\end{aligned}
 \end{equation}
Using these expressions, it is easy to write the large $\mu$ expansion (\ref{Jmu6pi}) of $J(\mu,6\pi)$ up to arbitrarily high order. 
 
Knowing the exact grand potential, we can also compute the spectral determinant and extract a quantization condition for the spectrum of $\rho$ (the fact that this is possible to do in the case $\hbar=6\pi$ was already anticipated in \cite{GHM}). Since $w_1$, $w_2$ and $f_1$ are invariant under $\mu \rightarrow \mu+2\pi {\rm i}$, we have
  \begin{equation}
 	J(\mu+2\pi {\rm i} n,6\pi)=J(\mu,6\pi)-{\rm i} \pi n^3+\left (-\frac{3}{2} \partial_t^2 F_0 \right )n^2+\frac{\rm i}{2\pi}\left ( \frac{t^2}{2}-\partial_t F_0^{\rm inst}+t \partial_t^2 F_0^{\rm inst}  -\frac{7\pi ^2}{6} \right )n.
 \end{equation}
 The spectral determinant defined in \cite{GHM} can be expressed using the Jacobi theta function $\theta_3(z;\tau)$ as
   \begin{equation}
 	\Xi(\mu,6\pi)= \sum_{n \in \mathbb Z} {\rm e}^{J(\mu+2\pi {\rm i} n,6\pi)}={\rm e}^{J(\mu,6\pi)} \theta_3(z;\tau),
 \end{equation}
 with
  \begin{equation}
  \begin{aligned}
 	z &=\frac{1}{4\pi^2}\left ( \frac{t^2}{6} -\partial_t F_0^{\rm inst}+t \partial_t^2 F_0^{\rm inst} \right )-\frac{19}{24}, \\
	\tau &=\frac{3 \rm i}{2 \pi}\partial_t^2 F_0.
\end{aligned}
 \end{equation}
 The spectral determinant vanishes when $\mu=E_n+{\rm i} \pi$, where the ``eigenenergies" $E_n$ are related to the eigenvalues $\lambda_n$ of $\rho$ as $\lambda_n={\rm e}^{-E_n}$.
 So we must look at the vanishing points of the theta function, which gives the quantization condition:
  \begin{equation}
  \begin{aligned}
 	z+\frac{\tau}{2}=n+\frac{1}{2},
\end{aligned}
 \end{equation}
 where $n$ is the integer label of $E_n$.
 Here are listed the first few numerical energies:
  \begin{equation}
 	\begin{array}{c}
	E_0 = 4.911789825 \\
	E_1 = 7.102439347 \\
	E_2 = 8.761483311 \\
	E_3 =10.15295695 \\
	E_4 = 11.37547713 \\
	E_5 = 12.47879902\\ 
	...\\
\end{array}
 \end{equation}
 It can be checked that the spectrum computed from the quantization condition numerically reproduces the traces (\ref{6pitraces}).


\end{document}